\documentclass[pagesize=pdftex,paper=a4,numbers=noenddot]{scrartcl}
\usepackage{geometry}
\usepackage{amsmath,amssymb,amstext,amsfonts}
\usepackage{mathrsfs}
\usepackage{bbm}
\usepackage{slashed}

\usepackage{lmodern}
\usepackage[T1]{fontenc}
\usepackage[english]{babel}
\usepackage[latin1]{inputenc}

\usepackage[pdftex]{graphicx}
\usepackage{caption,subcaption}
\usepackage{booktabs}
\usepackage{array}
\usepackage{xcolor} \definecolor{hellblau}{RGB}{70,126,185}
\usepackage[pdftex,linktocpage=true,pdfborder={0 0 0},colorlinks=true,linkcolor=hellblau,citecolor=hellblau,urlcolor=hellblau]{hyperref}
\usepackage[figuresright]{rotfloat}

\addtokomafont{captionlabel}{\small\bfseries}
\addtokomafont{caption}{\small}
\numberwithin{equation}{section}

\hypersetup{
pdftitle = {Relativistic SU(3) chiral baryon-baryon Lagrangian up to order p\^{}2},
pdfauthor = {Stefan Petschauer, Norbert Kaiser}
}
\title{Relativistic SU(3) chiral baryon-baryon Lagrangian up to order \boldmath $q^2$}
\author{Stefan Petschauer, Norbert Kaiser \\ Physik-Department, Technische Universität München, \\ D-85747 Garching, Germany \\ (September 2013)}
\date{}

\DeclareMathOperator{\diag}{diag}
\newcommand{\trace}[1]{\ensuremath{\langle #1\rangle}}
\newcommand{\Dl}{\ensuremath{\overleftarrow{D}}}
\newcommand{\thetal}{\ensuremath{\overleftarrow{\theta}}}
\newcommand{\xil}{\ensuremath{\overleftarrow{\xi}}}
\newcommand{\defeq}{\ensuremath{:=}}
\newcommand{\apr}{\ensuremath{\approx}}
\newcommand{\A}{\ensuremath{\hat{\mathcal{A}}}}

\newcommand{\C}{\ensuremath{\hat{\mathcal{C}}}}
\newcommand{\D}{\ensuremath{\hat{\mathcal{D}}}}
\newcommand{\E}{\ensuremath{\hat{\mathcal{E}}}}
\newcommand{\F}{\ensuremath{\hat{\mathcal{F}}}}
\newcommand{\extfield}{\,A\,}
\newcommand{\ho}{\ensuremath{\star}}
\newcommand{\ci}{\ensuremath{{c_i}}}

\begin{document}

\maketitle

\begin{quote}
\begin{center} \usekomafont{disposition} Abstract \end{center}
We construct the most general chiral effective Lagrangian for baryon-baryon contact interactions in flavor SU(3) up to order \(\mathcal O(q^2)\) using a covariant power counting.
A subset of these contact terms contributes to the baryon-baryon potential in chiral effective field theory.
The Lorentz invariant effective Lagrangian is constructed to fulfill the invariance under charge conjugation, parity transformation, Hermitian conjugation and the local chiral symmetry group \(\mathrm{SU}(3)_\mathrm L \times \mathrm{SU}(3)_\mathrm R\).
Goldstone bosons and external fields are included as well, thus providing additional four-baryon contact vertices involving e.g.\ pseudoscalar mesons and/or photons. In order to eliminate the linearly dependent terms, we use the Fierz identities, the equations of motion, and a Cayley-Hamilton relation for SU(3).
As an application the baryon-baryon scattering contact potentials in low partial waves are considered.
\end{quote}

\section{Introduction}

Since the seminal work of Weinberg \cite{Weinberg1968,Weinberg1978} chiral perturbation theory has become a powerful tool for calculating systematically the strong interaction dynamics in hadronic processes \cite{Gasser1984,Gasser1987,Bernard1995}.
This low-energy effective field theory is directly linked to the fundamental theory of strong interactions via chiral symmetry, its symmetry breaking patterns at low energies, and the discrete and Lorentz symmetries of quantum chromodynamics (QCD).
In combination with an appropriate power counting scheme, one can systematically improve the leading order calculation (equivalent to current algebra results) by including loop corrections and higher order Lagrangian terms.
The unresolved short-distance dynamics is encoded in contact terms, with a priori unknown low-energy constants (LECs).

At present a very accurate description of low-energy nucleon-nucleon scattering has been achieved in chiral effective field theory \cite{Kaiser1997,Kaiser1998,Epelbaum2004,Machleidt2011}.
The relatively large nucleon mass, \(M_N\approx 1 \text{ GeV}\), complicates the usual chiral power counting in small external momenta. This problem can, e.g., be overcome by using heavy baryon chiral perturbation theory \cite{Jenkins1990}, where the chiral Lagrangian is expanded in powers of the inverse baryon mass.
However there are some open issues in the Weinberg power counting scheme \cite{Kaplan1998b,Nogga2005}.
In the case of the \(NN\) potential one encounters at leading order two LECs and at next-to-leading order seven additional LECs.
The extension to the three-flavor case, relevant for describing the baryon-baryon interactions in all (strangeness and isospin) channels has not been treated in that detail, also due to the present shortage of experimental scattering data.
Hyperon masses in nuclear matter have been considered in \cite{Savage:1995kv}.
Hyperon masses in nuclear medium and hyperon-nucleon scattering were analyzed in \cite{Korpa2001} using an effective field theory in next-to-leading order.
The calculation of hyperon-nucleon scattering has been performed at leading order in chiral perturbation theory in \cite{Polinder2006,Haidenbauer:2007ra}.
This approximation provides already a good description of the available data.
Recently, a next-to-leading order calculation for hyperon-nucleon scattering has been presented in Ref.~\cite{Haidenbauer2013a}.
The chiral hyperon-nucleon potentials are also basic input for calculations of hypernuclei and strange baryonic matter.

The chiral Lagrangian for the baryon-number-one sector has been investigated in various works.
The two-flavor chiral effective pion-nucleon Lagrangian of order \(\mathcal O (q^4)\) has been constructed in Ref.~\cite{Fettes2000}.
The three-flavor Lorentz invariant chiral meson-baryon Lagrangians at order \(\mathcal O (q^2)\) and \(\mathcal O (q^3)\) have been first formulated in Ref.~\cite{Krause1990} and were later completed in Refs.~\cite{Oller2006} and \cite{Frink:2006hx}.
In these papers the external field method has been used and a locally chiral invariant relativistic Lagrangian has been derived.
Concerning the nucleon-nucleon contact terms, the general relativistically invariant two-flavor contact Lagrangian at order \(\mathcal O (q^2)\) (without any external fields) has been constructed in Ref.~\cite{Girlanda2010}.
In this paper we extend the framework and construct the general three-flavor baryon-baryon contact terms up to order \(\mathcal O(q^2)\) using the external field method.
The three-flavor chiral Lagrangian for the baryon-number-two sector is constructed such that each individual term fulfills the invariance under charge conjugation, parity transformation, Hermitian conjugation and local chiral symmetry \(\mathrm{SU}(3)_\mathrm L\times\mathrm{SU}(3)_\mathrm R\).
In order to eliminate linearly dependent terms, we use the Fierz identities, the equations of motion, the cyclic property of the trace and a Cayley-Hamilton relation for SU(3).
By employing the non-relativistic reduction we show that many of the relativistic \(\mathcal O(q^2)\)-terms actually contribute at higher order and thus can be discarded in non-relativistic calculations up to order \(\mathcal O(q^2)\).

\begin{figure}[htb]
 \centering
 \includegraphics[scale=0.8]{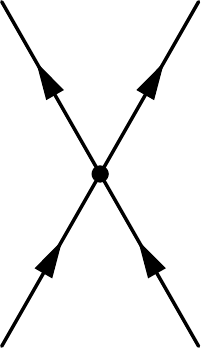}\qquad\qquad\qquad
 \includegraphics[scale=0.8]{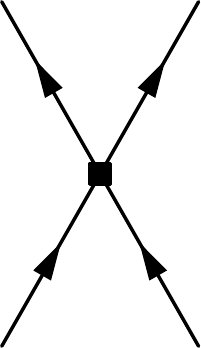}\qquad\qquad\qquad
 \includegraphics[scale=0.8]{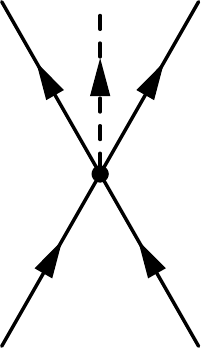}\qquad\qquad\qquad
 \includegraphics[scale=0.8]{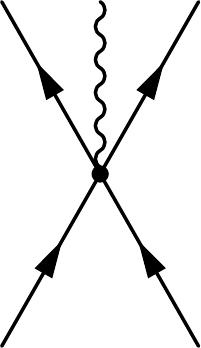}
 \caption{Examples for baryon-baryon contact vertices. The square symbolizes a higher order contact vertex. The dashed and wavy lines denote pseudoscalar mesons and photons, respectively.} \label{fig:ct}
\end{figure}

The resulting Lagrangian leads in the absence of external fields to the baryon-baryon contact terms up to \(\mathcal O(q^2)\), which are an important ingredient for complete calculations of the baryon-baryon interactions beyond leading order.
By introducing external fields of scalar, pseudoscalar, vector and axial vector type, the explicitly chiral symmetry breaking contact terms (linear in the quark mass) emerge as a particular subset.
Additional types of four-baryon contact vertices including pseudoscalar mesons and/or photons are obtained by this method as well, cf.\ Fig.~\ref{fig:ct}.
As one might expect, we find that the number of terms in the Lagrangian and associated low-energy constants is very large.
This is due to the fact, that many possible orderings of the flavor matrices including external fields give rise to invariant terms.
In addition, the particle content is considerably larger than for flavor SU(2).
However, many physical processes such as baryon-baryon scattering are sensitive to only a small fraction of these terms and their associated low-energy constants.

The present paper is organized as follows. In Sec.~\ref{sec:build} we introduce the basic building blocks for the chiral three-flavor Lagrangian and state their transformation properties. Sec.~\ref{sec:constr} is devoted to the construction principles that have been used to obtain the full set of chiral contact terms. In Sec.~\ref{sec:results} we present separately the expressions for the chiral Lagrangians of \(\mathcal O (q^0)\), \(\mathcal O (q^1)\) and \(\mathcal O (q^2)\).
These contact terms are evaluated for baryon-baryon scattering in Sec.~\ref{sec:application} and the corresponding potentials in a partial wave basis are shown.
The reduction of the number of terms gained by using the equation of motion is outlined in more detail in Appendix~\ref{sec:EOM}.
In Appendix~\ref{sec:HBexp} we present arguments which lead to a substantial reduction of the number of terms, when switching from the covariant power counting to the non-relativistic power counting.

\section{Building blocks of chiral Lagrangian} \label{sec:build}

In this section we summarize the basic procedure for constructing three-flavor chiral effective Lagrangians \cite{Coleman1969,Callan1969} with the inclusion of external fields \cite{Gasser1984,Gasser1985}.

Chiral perturbation theory is based on the QCD Lagrangian \(\mathscr L^0_\mathrm{QCD}\) with massless up, down and strange quarks \(u,d,s\).
It has a global chiral symmetry \(\mathrm{SU}(3)_\mathrm L \times \mathrm{SU}(3)_\mathrm R\), which is spontaneously broken to \(\mathrm{SU}(3)_\mathrm V\) by vacuum quark condensates.
In order to construct chirally invariant effective Lagrangians it is advantageous to introduce external fields, \(s(x)\), \(p(x)\), \(v_\mu(x)\), \(a_\mu(x)\),  of the form of Hermitian \(3\times3\) matrices that couple to scalar, pseudoscalar, vector and axial vector quark currents:
\begin{equation}
 \mathscr L = \mathscr L^0_\mathrm{QCD} + \bar q \gamma^\mu(v_\mu+\gamma_5a_\mu)q - \bar q(s-\mathrm i \gamma_5 p)q \,,\quad q=\begin{pmatrix} u\\d\\s \end{pmatrix} \,.
\end{equation}
By assuming a suitable transformation behavior of the external fields, the global chiral symmetry  \(\mathrm{SU}(3)_\mathrm L \times \mathrm{SU}(3)_\mathrm R\) gets promoted to a local gauge symmetry.
The non-vanishing current quark masses and in this way the explicit breaking of chiral symmetry can be introduced by setting the scalar field \(s(x)\) equal to the quark mass matrix, \(s(x)=\diag(m_u,m_d,m_s)\).
In the same way the electroweak interactions get included through appropriate external vector and axial vector fields.
In the case of the electromagnetic interaction one sets \(v_\mu(x)=eQA_\mu(x)\) with \(Q=\diag(2/3,-1/3,-1/3)\) the quark charge matrix  and \(A_\mu(x)\) the photon field.

The effective chiral Lagrangian has to fulfill all discrete and continuous symmetries of the strong interaction\footnote{We ignore the problem of strong CP violation.}. Therefore it has to be invariant under parity (P), charge conjugation (C), Hermitian conjugation (H) and the proper Lorentz transformations. Time reversal symmetry is then automatically fulfilled via the CPT theorem.
The explicit low-energy degrees of freedom of the chiral Lagrangian are the octet baryons (\(N, \Lambda, \Sigma, \Xi\)) and the pseudoscalar Goldstone-boson octet (\(\pi, K, \eta\)) in the standard non-linear realization of chiral symmetry \cite{Coleman1969,Callan1969}.

The chiral Lagrangian can be organized in the number of baryons,
\begin{equation*}
 \mathscr{L} = \mathscr{L}_\phi + \mathscr{L}_{MB} + \mathscr{L}_{BB} + \dots \,,
\end{equation*}
where \(\mathscr{L}_\phi\) is the purely mesonic part which has been constructed to \(\mathcal{O}(q^6)\) in Refs.~\cite{Fearing1994,Bijnens1999}. The meson-baryon interaction Lagrangian \(\mathscr{L}_{MB}\) has been derived to \(\mathcal{O}(q^3)\) in Refs.~\cite{Oller2006,Frink:2006hx}.
In this work we focus on the baryon-baryon contact Lagrangian \(\mathscr{L}_{BB}\) to \(\mathcal{O}(q^2)\). At lowest order \(\mathcal O (q^0)\) this Lagrangian has been given in Refs.~\cite{Savage:1995kv,Polinder2006}.

The building blocks of a chiral Lagrangian are composed of the external fields, \(s,p,v_\mu,a_\mu\), and the unitary matrix \(u^2=U = \exp\left(\mathrm i\phi(x)/f_0\right)\), with \(\phi(x)\) the pseudoscalar Goldstone-boson fields and \(f_0\) the meson decay constant in the chiral limit.
The particle content of \(\phi(x)\) and of the baryon matrix \(B(x)\) is
\begin{equation}
 \phi=
 \begin{pmatrix}
  \pi^0 + \frac{\eta}{\sqrt 3} & \sqrt 2 \pi^+ & \sqrt 2 K^+ \\
  \sqrt 2 \pi^- & -\pi^0 + \frac{\eta}{\sqrt 3} & \sqrt 2 K^0 \\
  \sqrt 2 K^- & \sqrt 2 \bar K^0 & -\frac{2\eta}{\sqrt 3}
 \end{pmatrix}
\ ,\quad
 B=
 \begin{pmatrix}
  \frac{\Sigma^0}{\sqrt 2} + \frac{\Lambda}{\sqrt 6} & \Sigma^+ & p \\
  \Sigma^- & -\frac{\Sigma^0}{\sqrt 2} + \frac{\Lambda}{\sqrt 6} & n \\
  \Xi^- & \Xi^0 & -\frac{2\Lambda}{\sqrt 6}
 \end{pmatrix}\,,
\end{equation}
where we have used the sign convention of \cite{Bernard1995} for meson and baryon fields.
With respect to the chiral transformation properties, the most convenient choice for the building blocks is given by
\begin{equation} \begin{aligned} \label{eq:buildingblocks}
 u_\mu &= \mathrm i \left[ u^\dagger\left(\partial_\mu-\mathrm i r_\mu \right) u - u\left(\partial_\mu - \mathrm i l_\mu\right)u^\dagger\right]\,,\\
 \chi_\pm &= u^\dagger\chi u^\dagger \pm u\chi^\dagger u\,,\\
 f_{\mu\nu}^\pm &= uf_{\mu\nu}^\mathrm L u^\dagger \pm u^\dagger f_{\mu\nu}^\mathrm R u\,,
\end{aligned} \end{equation}
with the combination
\begin{equation}
\chi= 2B_0\left(s+\mathrm ip\right)\,,
\end{equation}
of the external scalar and pseudoscalar field and the parameter \(B_0=-\langle0|\bar q q|0\rangle / 3f_0^2\), related to the vacuum quark condensate.
The external field strength tensors are defined by
\begin{equation}
 f^\mathrm R_{\mu\nu}= \partial_\mu r_\nu - \partial_\nu r_\mu - \mathrm i\left[r_\mu,r_\nu\right]\,,\qquad f^\mathrm L_{\mu\nu}= \partial_\mu l_\nu - \partial_\nu l_\mu - \mathrm i\left[l_\mu,l_\nu\right]\,, 
\end{equation}
where
\begin{equation}
 r_\mu=v_\mu+a_\mu\,,\quad l_\mu=v_\mu-a_\mu\,,
\end{equation}
denote right handed and left handed vector fields.
In the following we assume \(\trace{a_\mu}=\trace{v_\mu}=0\), where \(\trace{\dots}\) stands for the flavor trace. Hence the fields \(u_\mu\) and \(f^{\mu\nu}_\pm\) in Eq.~\eqref{eq:buildingblocks} are all traceless.

The Goldstone-boson fields \(U = \exp\left(\mathrm i\phi(x)/f_0\right)\) and the baryons fields \(B\) transform under the chiral symmetry group \(\mathrm{SU(3)}_\mathrm L\times \mathrm{SU(3)}_\mathrm R\) as follows \cite{Krause1990}
\begin{equation}
 \begin{pmatrix} U \\ B \end{pmatrix} \rightarrow \begin{pmatrix} RUL^\dagger \\ K(L,R,U) B K^\dagger(L,R,U) \end{pmatrix}\,,
\end{equation}
with \(L\in\mathrm{SU(3)}_\mathrm L\), \(R\in\mathrm{SU(3)}_\mathrm R\) and the SU(3)-valued compensator field \(K\left(L,R,U\right)\).
The square root \(u=\sqrt U\) transforms as \(u \rightarrow \sqrt{RUL^\dagger} = R u K^\dagger = K u L^\dagger\).

All building blocks \(A\), and therefore all products of these, transform according to the adjoint (octet) representation of SU(3), i.e.\ \(A\rightarrow KAK^\dagger\).
The chiral covariant derivative of such a field \(A\) is defined by
\begin{equation} \label{eq:covder}
 D_\mu A = \partial_\mu A + \left[\Gamma_\mu,A\right]\,,
\end{equation}
with the chiral connection
\begin{equation}
 \Gamma_\mu = \frac 1 2 \left[ u^\dagger \left(\partial_\mu -\mathrm i r_\mu\right) u + u \left(\partial_\mu -\mathrm il_\mu\right) u^\dagger \right]\,.
\end{equation}
The covariant derivative transforms homogeneously under the chiral group as \(D_\mu A \rightarrow K \left(D_\mu A\right) K^\dagger\).
The chiral covariant derivative of the baryon field \(B\) is given by Eq.~\eqref{eq:covder} as well.

We use the Lorentz-covariant power counting scheme, introduced by Krause in Ref.~\cite{Krause1990}.
Because of the large baryon mass \(M_0\) in the chiral limit, a time-derivative acting on a baryon field \(B\) cannot be counted as small. One has the following counting rules for baryon fields and their covariant derivatives,
\begin{equation}
 B\,,\ \bar B\,,\ D_\mu B \sim \mathcal{O}\left(q^0\right)\,,\qquad \left(\mathrm i \slashed{D} - M_0\right) B \sim \mathcal{O}\left(q\right)\,.
\end{equation}
The chiral dimension of the chiral building blocks and baryon bilinears \(\bar B\Gamma B\) are given in Table~\ref{tab:blocksbar}.
A covariant derivative acting on a building block (but not on \(B\)) raises the chiral dimension by one.

The transformation behavior of a building block \(A\) under parity, charge conjugation and Hermitian conjugation is
\begin{equation}
 A^P = (-1)^{p} A\,,\quad A^C = (-1)^{c} A^\top\,,\quad A^\dagger = (-1)^{h} A\,,
\end{equation}
with the exponents (modulo two) \(p,c,h\in\{0,1\}\) given in Table~\ref{tab:blocksbar}(\subref{subfig:blockbar}), and the transpose \(\top\) of a (flavor) matrix. In the case of parity \(P\), a sign change of the spatial argument, \(\left(t,\vec x\right) \rightarrow \left(t,-\vec x\right)\), is implied in the fields. Lorentz indices transform with the matrix \({P^\mu}_\nu=\diag(+1,-1,-1,-1)\) under the parity transformation, e.g., \((u^\mu)^P=(-1)^p{P^\mu}_\nu u^\nu\).
Commutators and anticommutators of two building blocks \(A_1,\ A_2\) have the same transformation behavior and therefore should be used instead of simple products, e.g.,
\begin{equation*}
 [A_1,A_2]_\pm^C = (-1)^{c_1+c_2}(A_1^\top A_2^\top \pm A_2^\top A_1^\top) = \pm (-1)^{c_1+c_2}[A_1,A_2]_\pm^\top\,.
\end{equation*}
For Hermitian conjugation the behavior is the same.
\begin{table}[tb]
\centering
\begin{subfigure}{.43\textwidth}
 \centering
 \begin{tabular}{>{\(}c<{\)}>{\(}c<{\)}>{\(}c<{\)}>{\(}c<{\)}>{\(}c<{\)}}
 \toprule
 & p & c & h & \mathcal{O} \\
 \cmidrule{1-1}\cmidrule(l){2-5}
 u_\mu & 1 & 0 & 0 & \mathcal{O}\left(q^1\right)\\
 f_{\mu\nu}^+ & 0 & 1 & 0 & \mathcal{O}\left(q^2\right)\\
 f_{\mu\nu}^- & 1 & 0 & 0 & \mathcal{O}\left(q^2\right)\\
 \chi_+ & 0 & 0 & 0 & \mathcal{O}\left(q^2\right)\\
 \chi_- & 1 & 0 & 1 & \mathcal{O}\left(q^2\right)\\
 \bottomrule
 \end{tabular}
 \caption{Chiral building blocks}\label{subfig:blockbar}
\end{subfigure}
\begin{subfigure}{.43\textwidth}
 \centering
 \begin{tabular}{>{\(}c<{\)}>{\(}c<{\)}>{\(}c<{\)}>{\(}c<{\)}>{\(}c<{\)}}
 \toprule
 \Gamma & p & c & h & \mathcal{O} \\
 \cmidrule{1-1}\cmidrule(l){2-5}
 \mathbbm{1} & 0 & 0 & 0 & \mathcal{O}\left(q^0\right)\\
 \gamma_5 & 1 & 0 & 1 & \mathcal{O}\left(q^1\right)\\
 \gamma_\mu & 0 & 1 & 0 & \mathcal{O}\left(q^0\right)\\
 \gamma_5\gamma_\mu & 1 & 0 & 0 & \mathcal{O}\left(q^0\right)\\
 \sigma_{\mu\nu} & 0 & 1 & 0 & \mathcal{O}\left(q^0\right)\\
 \bottomrule
 \end{tabular}
 \caption{Baryon bilinears \(\bar B \Gamma B\)}\label{subfig:clifford}
\end{subfigure}
\caption{Behavior under parity, charge conjugation and Hermitian conjugation as well as the chiral  dimensions of chiral building blocks and baryon bilinears \(\bar B \Gamma B\) \cite{Oller2006}.} 
\label{tab:blocksbar}
\end{table}
The basis elements of the Dirac algebra forming the baryon bilinears have the transformation behavior
\begin{equation}
 \gamma_0\Gamma\gamma_0 = (-1)^{p_\Gamma} \Gamma\,,\quad C^{-1}\Gamma C = (-1)^{c_\Gamma} \Gamma^\top\,,\quad \gamma_0\Gamma^\dagger\gamma_0 = (-1)^{h_\Gamma} \Gamma\,,
\end{equation} 
where the exponents \(p_\Gamma, c_\Gamma, h_\Gamma\in\{0,1\}\) can be found in Table~\ref{tab:blocksbar}(\subref{subfig:clifford}). Again, Lorentz indices of baryon bilinears transform under parity with the matrix \({P^\mu}_\nu\).

Because of the relation
\begin{equation}
\left[D_\mu,D_\nu\right] A = \frac14 \left[\left[u_\mu,u_\nu\right],A\right] - \frac{\mathrm i}{2} \left[f^+_{\mu\nu},A\right]
\end{equation}
for any building block \(A\) (or baryon field \(B\)), it is sufficient to use only totally symmetrized products of covariant derivatives, \(D^{\alpha\beta\gamma\dots}A\).
Furthermore, because of the identity
\begin{equation}
 D_\nu u_\mu - D_\mu u_\nu = f^-_{\mu\nu}\,,
\end{equation}
one needs to consider only the symmetrized covariant derivative acting on \(u_\mu\,\),
\begin{equation} \label{eq:defhmunu}
 h_{\mu\nu}\defeq D_\mu u_\nu + D_\nu u_\mu\,.
\end{equation}

\section{Construction of chiral Lagrangian} \label{sec:constr}

For the construction of all terms in the Lagrangian, invariance under local chiral transformations, proper Lorentz transformations, parity, charge conjugation and Hermitian conjugation has to be fulfilled.
Since the baryon field \(B\) and all building blocks \(A\) transform under the chiral symmetry group in the same way, \(A\rightarrow KAK^\dagger\), the invariant terms for the Lagrangians are constructed by traces over products of these, or by products of such traces. Note that \(K^\dagger K=\mathbbm{1}\).
The baryon-baryon contact terms include two baryon fields \(B\) and two adjoint baryon fields \(\bar B\).
The different arrangements of these four baryon fields (exploiting the cyclic property of the trace) are of the schematic form:
\begin{gather*}
 \trace{\bar B B \bar B B}\,,\ \trace{\bar B \bar B B B}\,,\\
 \trace{\bar B B}\trace{\bar B B}\,,\ \trace{\bar B}\trace{B \bar B B}\,,\ \trace{\bar B \bar B}\trace{BB}\,,\ \trace{\bar B B \bar B}\trace{B} \,,\\
 \trace{\bar B B}\trace{\bar B}\trace{B}\,,\ \trace{\bar B}\trace{\bar B}\trace{B B}\,,\ \trace{\bar B\bar B}\trace{B}\trace{B}\,,\\
 \trace{\bar B}\trace{B} \trace{\bar B}\trace{B}\,.
\end{gather*}
Chiral building blocks and Dirac operators have to be supplemented appropriately.
By using the Fierz identity, a product of two baryon bilinears can be rearranged
\begin{equation}
 (\bar \Psi^{(1)}\Gamma^A\Psi^{(2)})(\bar \Psi^{(3)}\Gamma^B\Psi^{(4)}) = -\sum_{C,D} \mathcal C ^{AB}_{CD} (\bar \Psi^{(1)}\Gamma^C\Psi^{(4)})(\bar \Psi^{(3)}\Gamma^D\Psi^{(2)}) \,,
\end{equation}
where \(\mathcal C^{AB}_{CD}\) are the well-known Fierz transformation coefficients.
This allows one to fix and label the fields \(\bar B\) and \(B\) which form the first and second baryon bilinear, respectively.
The other arrangement can be expressed as a linear combination of the chosen type.

One arrives at the following list of general flavor structures for contact terms to arbitrary chiral order:
\begin{equation} \begin{aligned}\label{eq:defX}
X^1 &\defeq \hat D_2^k \, \trace{ \bar B_1 A_a \Theta_1 B_1 A_b \bar B_2 A_c\Theta_2 B_2 A_d },
\\
X^2 &\defeq \hat D_2^k \, \trace{ \bar B_1 A_a \bar B_2 A_b \Theta_1 B_1 A_c\Theta_2 B_2 A_d },
\\
X^3 &\defeq \hat D_2^k \, \trace{ \bar B_1 A_a \Theta_1 B_1 A_b } \, \trace{ \bar B_2 A_c\Theta_2 B_2 A_d },
\\
X^4 &\defeq \hat D_2^k \, \trace{ \bar B_1 A_a } \, \trace{ \Theta_1 B_1 A_b \bar B_2 A_c\Theta_2 B_2 A_d },
\\
X^5 &\defeq \hat D_2^k \, \trace{ \bar B_1 A_a \Theta_1 B_1 A_b } \, \trace{ \bar B_2 A_c} \, \trace{\Theta_2 B_2 A_d},
\\
X^6 &\defeq \hat D_2^k \, \trace{ \bar B_1 A_a} \, \trace{ \bar B_2 A_b } \, \trace{ \Theta_1 B_1 A_c\Theta_2 B_2 A_d },
\\
X^7 &\defeq \hat D_2^k \, \trace{ \bar B_1 A_a} \, \trace{ \Theta_1 B_1 A_b } \, \trace{ \bar B_2 A_c} \, \trace{\Theta_2 B_2 A_d }, \\
&\text{and all terms of the form: } X^i\cdot\langle A_e\rangle\cdot \langle A_f\rangle\,\dots \,,
\end{aligned} \end{equation}
where the \(A\) are either \(\mathbbm1\) or the building blocks (\(u_\mu,\ f_{\mu\nu}^\pm,\ \chi_\pm\)), covariant derivatives thereof, and commutators or anticommutators of these. We have omitted their Lorentz indices for the sake of notational simplicity.
The operators \(\Theta_{1,2}\) consist of basis elements of the Dirac algebra (in order to get the complete set of baryon bilinears) and products of the metric tensor \(g_{\mu\nu}\) and the Levi-Civita tensor \(\epsilon_{\mu\nu\rho\lambda}\).
They include also an arbitrary number (\(n_{1,2}\)) of totally symmetrized covariant derivatives acting on the baryon field \(B\) to the right: \(\Theta_1 \defeq \Gamma_1 D^{n_1}\,,\ \Theta_2 \defeq \Gamma_2 D^{n_2}\,,\ \Gamma_i \in \{\mathbbm{1},\gamma_5, \gamma_\mu, \gamma_5\gamma_\mu, \sigma_{\mu\nu}\}\).
The indices 1,2 of the baryon fields and \(\Theta\)'s indicate to which baryon bilinear they belong, e.g.\ (\(\dots\bar B_1\dots\Theta_1B_1\dots\)) means (\(\dots\bar B_\alpha\dots\Theta^{\alpha\beta}B_\beta\dots\)) with \(\alpha,\ \beta\) being spinor indices.
The operator \(\hat D_2^k\) is defined such, that it acts only on the baryon field indexed by 2, e.g.\ \(\hat D_2(\dots\bar B_2\dots B_2\dots)\) means \((\dots(D\bar B_2)\dots B_2\dots) + (\dots\bar B_2\dots (DB_2)\dots)\). Furthermore, these \(k\) covariant derivatives are totally symmetrized.
Each operator \(\hat D_2\) raises the chiral power of the monomial by one.

Obviously, a total derivative term \(\partial X\) can be omitted from the Lagrangian. In our case this gives, by the use of the product rule\footnote{The product rule reads \(D^\mu(A_aA_b) = A_a(D^\mu A_b) + (A_a\Dl^\mu)A_b\) with \(A\Dl^\mu \defeq D^\mu A\). We also use the relation \(\partial_\mu\trace{\dots} = \trace{D_\mu(\dots)}\) for trace terms.},
\begin{equation} \label{eq:totalder}
 \partial X = D X = \hat D_1 X + \hat D_2 X + \sum(\dots DA\dots)\,,
\end{equation}
where \((\dots DA\dots)\) denotes a term including the covariant derivatives of a chiral building block.
Since \(DA\) is by definition a chiral building block, the corresponding term is already included in the construction of the most general Lagrangian.
As a consequence of Eq.~\eqref{eq:totalder} there is no need to consider the operator \(\hat D_1^k\) since it can be replaced by \(-\hat D_2^k\).
In a similar way we do not need to consider covariant derivatives of \(\bar B\), since these can be expressed by higher order terms and terms that are already included in the list \(X^1,\dots,X^7\) of Eq.~\eqref{eq:defX}. This property follows from the definition of \(\hat D_2\),
\begin{equation} \label{eq:elimBb}
 \hat D_2 (\dots \bar B_2\dots B_2\dots) = (\dots (D\bar B_2)\dots B_2 \dots) + (\dots \bar B_2 \dots (DB_2)\dots) \,.
\end{equation} 

The arrangement \(\trace{\bar B \bar B}\trace{BB}\) of baryon fields under the flavor traces, which stands for a general term \(\hat D_2^k \, \trace{ \bar B_1 A_a \bar B_2 A_b } \, \trace{ \Theta_1 B_1 A_c\Theta_2 B_2 A_d }\), can be expressed by other arrangements, using the SU(3) Cayley-Hamilton relation, cf.\ Refs.~\cite{Frink:2006hx,Bijnens1999,Zhang2007}, together with the Fierz identity,
\begin{equation} \begin{aligned}
 0 =& \sum_\text{6 perm} \trace{M_1M_2M_3M_4} - \sum_\text{8 perm} \trace{M_1M_2M_3}\trace{M_4} - \sum_\text{3 perm} \trace{M_1M_2}\trace{M_3M_4} \\
&+ \sum_\text{6 perm} \trace{M_1M_2}\trace{M_3}\trace{M_4} - \trace{M_1}\trace{M_2}\trace{M_3}\trace{M_4}\,,
\end{aligned} \end{equation}
with \(M_1=\bar B_1 A_a \,,\ M_2=\bar B_2 A_b \,,\ M_3=\Theta_1 B_1 A_c \,,\ M_4=\Theta_2 B_2 A_d \).
The arrangements \(\trace{\bar B \bar B B}\trace{B}\) and \(\trace{\bar B\bar B}\trace{B}\trace{B}\) arise by charge conjugation from the monomials \(X^4\) and \(X^6\) and therefore do not need to be included explicitly.

\subsection*{Transformation behavior of monomials \boldmath{\(X^i\)} under parity}

Each monomial \(X^i\) in the list of Eq.~\eqref{eq:defX} transforms under parity as
\begin{equation} \label{eq:paritytrans}
 (X^i)^P = (-1)^{p_A+p_{\Gamma_1}+p_{\Gamma_2}+n_\epsilon} X^i = (-1)^{p} X^i\,,
\end{equation} 
where \(p_A\) is the sum of parity exponents of the external fields \(p_A = \sum_{j=a,b,c,d,\dots} p_{A_j}\) (cf. Table~\ref{tab:blocksbar}(\subref{subfig:blockbar})).
Likewise, \(p_{\Gamma_i}\) is the parity exponent of the Dirac algebra element in \(\Theta_i\) (cf. Table~\ref{tab:blocksbar}(\subref{subfig:clifford})), deduced from the transformation \(\bar\Psi\Gamma\Phi\overset{P}{\rightarrow}\bar\Psi\gamma_0\Gamma\gamma_0\Phi\), and \(n_\epsilon\) counts the number of Levi-Civita tensors in the monomial \(X^i\).
The counting rule for the sign in Eq.~\eqref{eq:paritytrans} holds, because all Lorentz indices of the building blocks and also of the Dirac algebra basis elements transform as \( I_{\mu}^P = \pm {P^\nu}_\mu I_\nu \) with \(({P^\nu}_\mu)=\diag(+1,-1,-1,-1)\) and therefore\footnote{Covariant derivatives transform as four-vectors under parity, \((D^\mu A)^P = {P^\mu}_\nu (D^\nu A^P)\).}
\begin{equation}
 I_{\mu}^P J^{\mu,P} = \pm I_\rho {P^\rho}_\mu {P^\mu}_\nu J^\nu = \pm I_{\mu} J^{\mu}\,.
\end{equation}
With inclusion of the Lorentz indices we have the tensorial structures \(A_{\mu(\nu)},D_\mu,\Gamma_{\mu(\nu)},g_{\mu\nu},\epsilon_{\mu\nu\rho\sigma}\), where the metric \(g_{\mu\nu}\) can be dropped, since it merely raises or lowers indices.
Because of the relation
\begin{equation}
 P^{\mu}_{\ \mu^\prime}P^{\nu}_{\ \nu^\prime}P^{\rho}_{\ \rho^\prime}P^{\sigma}_{\ \sigma^\prime}\epsilon^{\mu^\prime\nu^\prime\rho^\prime\sigma^\prime}   = \det(P)\epsilon^{\mu\nu\rho\sigma} = - \epsilon^{\mu\nu\rho\sigma} \,,
\end{equation}
one gets a minus sign for each \(\epsilon\)-tensor that appears in a monomial term.

\subsection*{Transformation behavior of monomials \boldmath{\(X^i\)} under charge conjugation}

Here, we analyze the transformation behavior of the monomials \(X^i\) listed in Eq.~\eqref{eq:defX} under charge conjugation.
The symbol \(\top\) denotes transposition of matrices in Dirac space as well as in flavor space.
For the first monomial \(X^1\) one gets
\begin{equation}\begin{aligned}
 (X^1)^C &= \left(\hat D_2^k \, \trace{ \bar B_1 A_a \Theta_1 B_1 A_b \bar B_2 A_c\Theta_2 B_2 A_d }\right)^C \\
 &= (-1)^{c_A+c_{\Gamma_1}+c_{\Gamma_2}}\hat D_2^k \, \trace{ B_1^{\top} A_a^{\top} (\bar B_1 \overleftarrow{\Theta}_1 )^{\top} A_b^{\top} B_2^{\top} A_c^{\top}(\bar B_2 \overleftarrow{\Theta}_2)^{\top} A_d^{\top} } \\
 &= (-1)^{c_A+c_{\Gamma_1}+c_{\Gamma_2}}\hat D_2^k \, \trace{ A_d \bar B_2 \overleftarrow{\Theta}_2 A_c B_2 A_b \bar B_1 \overleftarrow{\Theta}_1 A_a B_1}\\
 &= (-1)^{c}\hat D_2^k \, \trace{\bar B_1 \overleftarrow{\Theta}_1 A_a B_1 A_d \bar B_2 \overleftarrow{\Theta}_2 A_c B_2 A_b}\\
 &= (-1)^{c+n_1+n_2} X^1_{b \leftrightarrow d} + \text{h.o.}\,,
\end{aligned}\end{equation}
where we have used in the first step \(B^C=C \bar B^{\top}\) and \(\bar B^C=-B^{\top}C^{-1}\), with \(C=\mathrm i \gamma^2\gamma^0\).
The relations \(C^{-1}\Gamma_iC=(-1)^{c_{\Gamma_i}}\Gamma_i^\top\), \(A_i^C=(-1)^{c_i} A_i^\top\) and \(D_\mu^C A^{\top} = (D_\mu A)^{\top}\) give the charge conjugation for Dirac matrices, external fields and covariant derivatives.
One obtains the exponents \(c_A = \sum_{j=a,b,c,d,\dots} c_{A_j}\) and \(c= c_A+c_{\Gamma_1}+c_{\Gamma_2}\in\{0,1\}\).
In the step from the second to the third line one uses  the relation \(\Psi^{\top} \Gamma^{\top} \bar\Phi^{\top} = \Psi_\alpha \Gamma_{\beta\alpha} \bar\Phi_\beta = -\bar\Phi\Gamma\Psi\) for bilinears in Dirac space and the relation \((A^{\top} B^{\top} C^{\top}\dots)=(\dots CBA)^{\top}\) for flavor matrices, together with \(\trace{A^{\top}}=\trace{A}\).
The notation \(\overleftarrow{\Theta}_i = \Dl^{n_i} \Gamma_i\) indicates that the covariant derivatives act on the baryon field to the left.
For the last equality we used the product rule Eq.~\eqref{eq:elimBb} and the abbreviation h.o.\ denotes higher order terms.

Similarly, one finds for the monomials \(X^2,X^3\) and \(X^4\):
\begin{align}
 (X^2)^C
 = (-1)^{c} \hat D_2^k \, \trace{ \bar B_2 \overleftarrow\Theta_2 A_c \bar B_1 \overleftarrow\Theta_1 A_b B_2 A_a B_1 A_d }
 = (-1)^{c+n_1+n_2+k} X^2_{\substack{a\leftrightarrow c \\ \Gamma_1\leftrightarrow\Gamma_2 \\ n_1\leftrightarrow n_2}} + \text{h.o.}\,,
\end{align}
\begin{align}
 (X^3)^C
  = (-1)^{c} \hat D_2^k \, \trace{ \bar B_1 \overleftarrow\Theta_1 A_a B_1 A_b } \, \trace{ \bar B_2 \overleftarrow\Theta_2 A_c B_2 A_d }
  = (-1)^{c+n_1+n_2} X^3 + \text{h.o.}\,,
\end{align} 
\begin{align}
 (X^4)^C
  = (-1)^{c} \hat D_2^k \, \trace{ \bar B_2 \overleftarrow\Theta_2A_cB_2A_b\bar B_1\overleftarrow\Theta_1 A_d} \, \trace{ B_1 A_a }\,.
\end{align} 
At this point one can see that every term of the schematic form \(\trace{\bar B B \bar B}\trace{B}\) can be written as the charge-conjugate of a term of the form \(\trace{\bar B}\trace{B \bar B B}\) (and vice versa).
Since only charge conjugation invariant terms \(X+X^C\) are allowed, it is sufficient to consider only the form \(\trace{\bar B}\trace{B \bar B B}\), i.e.\ terms of the type \(X^4\).
We continue with the charge conjugation properties of the monomials \(X^5\) and \(X^6\):
\begin{align}
 (X^5)^C
  = (-1)^{c} \hat D_2^k \, \trace{ \bar B_1 \overleftarrow\Theta_1 A_a B_1 A_b } \, \trace{ \bar B_2 \overleftarrow\Theta_2 A_d} \, \trace{B_2 A_c}
  = (-1)^{c+n_1+n_2} X^5_{\substack{c\leftrightarrow d}} + \text{h.o.}\,,
\end{align} 
\begin{align}
 (X^6)^C
  = (-1)^{c} \hat D_2^k \, \trace{\bar B_2 \overleftarrow\Theta_2 A_c \bar B_1 \overleftarrow\Theta_1 A_d}\, \trace{B_2 A_b} \, \trace{B_1 A_a}\,.
\end{align}
As before, the schematic forms \(\trace{\bar B\bar B}\trace{B}\trace{B}\) and \(\trace{\bar B}\trace{\bar B}\trace{B B}\) are connected by charge conjugation. Therefore it is sufficient to consider only \(\trace{\bar B\bar B}\trace{B}\trace{B}\), i.e.\ terms of the type \(X^6\).
Finally, the charge conjugation property of \(X^7\) is
\begin{align}
 (X^7)^C
  = (-1)^{c} \hat D_2^k \, \trace{ \bar B_1 \overleftarrow\Theta_1 A_b} \, \trace{ B_1 A_a } \, \trace{ \bar B_2 \overleftarrow\Theta_2 A_d} \, \trace{ B_2 A_c }
  = (-1)^{c+n_1+n_2} X^7_{\substack{a\leftrightarrow b\\c\leftrightarrow d}} + \text{h.o.}\,.
\end{align}
For terms with additional traces multiplied to \(X^i\) (see Eq.~\eqref{eq:defX}) the behavior under charge conjugation follows as
\begin{equation}
 (X^i\cdot\trace{A_e}\cdot\trace{A_f}\cdot\dots)^C = (-1)^{c_e+c_f+\dots}(X^i)^C\cdot\trace{A_e}\cdot\trace{A_f}\cdot\dots\,.
\end{equation}

\subsection*{Transformation behavior of monomials \boldmath{\(X^i\)} under Hermitian conjugation}

Now we consider the transformation behavior of the monomials \(X^i\) listed in Eq.~\eqref{eq:defX} under Hermitian conjugation.
For the first monomial \(X^1\) we have
\begin{equation}\begin{aligned}
 (X^1)^* &= \hat D_2^k \, \trace{ \bar B_1 A_a \Theta_1 B_1 A_b \bar B_2 A_c\Theta_2 B_2 A_d }^* \\
 &= \hat D_2^k \, \trace{ A_d^\dagger B_2^\dagger \overleftarrow{\Theta}_2^\dagger A_c^\dagger \bar B_2^\dagger A_b^\dagger B_1^\dagger \overleftarrow{\Theta}_1^\dagger A_a^\dagger \bar B_1^\dagger}\\
 &= (-1)^{h_A+h_{\Gamma_1}+h_{\Gamma_2}} \hat D_2^k \, \trace{ A_d \bar B_2 \overleftarrow{\Theta}_2 A_c B_2 A_b \bar B_1 \overleftarrow{\Theta}_1 A_a B_1}\\
 &= (-1)^{h} \hat D_2^k \, \trace{\bar B_1 \overleftarrow{\Theta}_1 A_a B_1 A_d \bar B_2 \overleftarrow{\Theta}_2 A_c B_2 A_b}\\
 &= (-1)^{h+n_1+n_2} X^1_{b \leftrightarrow d} + \text{h.o.} \,,
\end{aligned}\end{equation}
where we used in the first step the relations \(\trace{A^*}=\trace{A^\dagger}\), \((ABC\dots)^\dagger=(\dots C^\dagger B^\dagger A^\dagger)\) for flavor matrices and the relation \((\bar \Psi \Gamma\Phi)^* = (\bar\Psi \Gamma\Phi)^\dagger = \Phi^\dagger \Gamma^\dagger\bar\Psi^\dagger\) for (mixed) baryon bilinears.
Covariant derivatives follow the rule \((D^\mu A)^\dagger = A^\dagger\Dl^\mu\), and the notation \(\overleftarrow{\Theta}_i^\dagger = \Dl^{n_i} \Gamma_i^\dagger\) means that the Hermitian conjugate of \(\Theta_i\) acts only on the Dirac matrix.
In the second step we used the properties \(A_i^\dagger = (-1)^{h_i} A_i\) and \(\gamma_0 \Gamma_i^\dagger\gamma_0 = (-1)^{h_{\Gamma_i}}\Gamma_i\).
One obtains the exponents \(h_A = \sum_{j=a,b,c,d,\dots} h_{A_j}\) and \(h = h_A+h_{\Gamma_1}+h_{\Gamma_2}\in\{0,1\}\).
In the last step the product rule Eq.~\eqref{eq:elimBb} has been employed.

Just as for the monomial \(X^1\), the transformation behavior under Hermitian conjugation of the other monomials \(X^i\) is given by the replacement of the exponents \(c\rightarrow h\) in the transformation under charge conjugation.

\section{Results for chiral contact terms} \label{sec:results}

In this section we apply the rules for constructing a chiral Lagrangian to obtain a complete set of three-flavor contact terms up to order \(\mathcal O (q^2)\).
We define for an arbitrary monomial \(X\) the charge conjugation invariant combination \(Y=X+X^C\).
It transforms under parity as \((Y)^{P} = (-1)^{p} Y\), since \(X\) and \(X^C\) transform into themselves up to a sign factor \((-1)^{p_X}\).
The behavior under Hermitian conjugation is \(Y^{*} = (-1)^{c+h} Y\), since
\begin{equation} \label{eq:defXarrow}
 (X+X^C)^* = X^*+X^{C*}
= (-1)^{c+h} (X^C + X)\,.
\end{equation}
Here the similar transformation behavior under charge conjugation and Hermitian conjugation led to the relation \(X^*=(-1)^{c+h} X^C\).
If \(Y\) transforms under Hermitian conjugation into its negative, one has to multiply it with a factor \(\mathrm i\).
In the cases where \(X^{C} = - X + \text{h.o.}\) one can drop these monomials, since \(Y\) is then zero to the considered order.

At a given order there are arbitrarily many terms with pairwise contracted covariant derivatives of the form
\begin{equation} \label{eq:argument}
  Y(\Theta_1 = \ldots \cdot D_{\mu_1\mu_2\dots\mu_n}\,,\ \Theta_2 = \ldots \cdot D^{\mu_1\mu_2\dots\mu_n})\,.
\end{equation}
Following an argument of Ref.~\cite{Girlanda2010}, one needs to take into account values for \(n\) only up to a finite number for the Lagrangian to order \(\mathcal{O}(q^2)\).
A term of the form in Eq.~\eqref{eq:argument} gives rise to a matrix element \((\bar u_3\Theta_1 u_1)(\bar u_4 \Theta_2 u_2)\),
where every contracted pair of \(D\)'s produces a factor \(p_1 \cdot p_{2}\).
Up to \(\mathcal{O}(q^2)\) one can approximate its \(n\)-th power as
\begin{equation}
 (p_1 \cdot p_2)^n \approx (M_0)^{2n}\left[1+\frac{n}{2M_0^2}(\vec p_1-\vec p_2)^2\right]\,,
\end{equation}
and therefore all \(n\) larger than 1 give not rise to new structures.
Because of the field \(\Gamma_\mu\) (which is of \(\mathcal O(q)\)) in the covariant derivative one needs to go one order higher.
At \(\mathcal O(q^0)\) terms with \(n=0,1,2\) are needed, at \(\mathcal O(q)\) terms with \(n=0,1\) can contribute and at \(\mathcal O(q^2)\) only terms with \(n=0\) need to be considered.

After a non-relativistic reduction one finds that if a \(D_\mu\) acting on a baryon field \(B\) is contracted with one of the Dirac matrices \(\{\gamma_5\gamma_\mu,\sigma_{\mu\nu}\}\), the term actually contributes at least one order higher, proportional to \(q/M_0\).
The corresponding expansions are given in Appendix~\ref{sec:HBexp}.
In the following tables we will label such suppressed terms by an asterisk \(\ho\,\).
It is still possible that independent terms in the covariant power counting become equal in the non-relativistic power counting, even though none of them is suppressed.
This leads to a further reduction of independent terms and associated low-energy constants.

We give now the possible terms for the relativistic contact Lagrangian up to order \(\mathcal{O}(q^2)\).
By the use of the lowest order equations of motion satisfied by baryons,
\begin{equation}
 \left(\mathrm i\slashed{D} - M_0\right) B = 0\,,
\end{equation} 
the number of independent terms can be reduced considerably.
The details of this reduction are worked out in Appendix~\ref{sec:EOM}.
Furthermore, by using the lowest order equation of motion satisfied by the mesons,
\begin{equation}
 D_\mu u^\mu = \frac{\mathrm i}{2} \left(\chi_- - \frac13\langle\chi_-\rangle \right)\,,
\end{equation} 
one can get rid of terms including \({h_\mu}^\mu = D_\mu u^\mu + D^\mu u_\mu\) (see Eq.~\eqref{eq:defhmunu}).

\subsection[Contact terms of \texorpdfstring{$\mathcal{O}(q^0)$}{O(q\^{}0)}]{Contact terms of \boldmath{$\mathcal{O}(q^0)$}} \label{subsec:results-q0}

At order \(\mathcal{O}(q^0)\) only terms of the type \(X^{1,2,3}\) contribute, since there are no external fields present, and \(\trace{B}=\trace{\bar B}=0\).
The leading order three-flavor contact Lagrangian has 15 terms and is given by
\begin{equation}
 \mathscr{L}^{(0)}_{BB} = \sum_{i=1}^{5}  \left(a_{1,i} \A^1_i + a_{2,i} \A^2_i +  a_{3,i} \A^3_i\right) \,,
\end{equation}
with the flavor structures
\begin{equation} \begin{aligned} \label{eq:loflavor}
 \A^1_i &= \trace{\bar B_1\theta^iB_1\bar B_2\xi^iB_2} + \trace{\bar B_1\thetal^iB_1\bar B_2\xil^iB_2}\,,\\
 \A^2_i &= \trace{\bar B_1\bar B_2\theta^iB_1\xi^iB_2} + \trace{\bar B_1\xil^i\bar B_2\thetal^iB_1B_2}\,,\\
 \A^3_i &= \trace{\bar B_1\theta^iB_1}\trace{\bar B_2\xi^iB_2} + \trace{\bar B_1\thetal^iB_1}\trace{\bar B_2\xil^iB_2}\,,
\end{aligned} \end{equation}
and the operators \(\theta^i\), \(\xi^i\) given in Table~\ref{tab:ct0}. The real parameters \(a_{j,i}\) are the associated low-energy constants.
As in Ref.~\cite{Oller2006} we choose the terms to be exactly invariant under charge and Hermitian conjugation, and not just invariant to leading order.
Therefore both summands in Eq.~\eqref{eq:loflavor} are needed.
As stated above additionally the Dirac operators in Table~\ref{tab:ct0} with one contracted pair of covariant derivatives (e.g.\ \(\sigma^{\mu\nu}D^\rho \otimes \sigma_{\mu\nu}D_\rho\)) and with two contracted pairs (e.g.\ \(\sigma^{\mu\nu}D^{\rho\tau} \otimes \sigma_{\mu\nu}D_{\rho\tau}\)) have to be included.

Considering this Lagrangian in the non-relativistic approximation for baryon-baryon scattering, we recover the results of Ref.~\cite{Polinder2006}.
Dirac operators such as \(\sigma^{\mu\nu}\partial^\rho \otimes \sigma_{\mu\nu}\partial_\rho\) give in leading order the same contribution as the ones without contracted derivatives, but differ at higher order.
To leading order in the non-relativistic expansion the only contributions come from the Dirac operators \(\mathbbm{1} \otimes \mathbbm{1}\) and \(\gamma_5\gamma^\mu \otimes \gamma_5\gamma_\mu\).
The others are either of higher order or give contributions equal to these two. As a result one has six  independent non-relativistic contact terms at leading order.
This is consistent with group theoretical considerations, were the product of two (baryon) octets is decomposed into a sum of six irreducible SU(3) representations, \( \mathbf8\otimes\mathbf8 = \mathbf{27}_s\oplus\mathbf{10}_a\oplus\mathbf{10^*}_a\oplus\mathbf{8}_s\oplus\mathbf{8}_a\oplus\mathbf1_s \).
The symmetric and antisymmetric flavor representation are combined with the spin singlet and spin triplet states, respectively.
The parameters of the leading order Lagrangian \(\mathscr{L}^{(0)}_{BB}\) can be combined to the low-energy constants for irreducible SU(3) representations and spin multiplets, which are used also later in Table~\ref{tab:PWDBB}. The corresponding relations read:
\begin{equation}\begin{aligned}
 \tilde c_{\,{}^1\!S_0}^{27}&=8\pi\left[2 (a_{1,1}+3a_{1,2})+2 (a_{3,1}+3a_{3,2})\right]\,, \\
 \tilde c_{\,{}^1\!S_0}^{8s}&=8\pi\left[-\frac{4}{3}(a_{1,1}+3a_{1,2})-\frac{5}{3}(a_{2,1}+3a_{2,2})+2 (a_{3,1}+3a_{3,2})\right]\,, \\
 \tilde c_{\,{}^1\!S_0}^{1}&=8\pi\left[-\frac{2}{3}(a_{1,1}+3a_{1,2})-\frac{16}{3}(a_{2,1}+3a_{2,2})+2 (a_{3,1}+3a_{3,2})\right]\,, \\
 \tilde c_{\,{}^3\!S_1}^{8a}&=8\pi\left[3 (a_{2,1}-a_{2,2})+2 (a_{3,1}-a_{3,2})\right]\,, \\
 \tilde c_{\,{}^3\!S_1}^{10}&=8\pi\left[-2 (a_{1,1}-a_{1,2})+2 (a_{3,1}-a_{3,2})\right]\,, \\
 \tilde c_{\,{}^3\!S_1}^{10^*}&=8\pi\left[2 (a_{1,1}-a_{1,2})+2 (a_{3,1}-a_{3,2})\right] \,.
\end{aligned}\end{equation}

\begin{table}[tb]
\centering
\begin{tabular}{>{\(}c<{\)}>{\(}c<{\)}>{\(}c<{\)}>{\(}c<{\)}}
\toprule
 \theta^i & \xi^i & \text{NR} & \text{contributes to } \A^j\\
\midrule
 \mathbbm{1} & \mathbbm{1} & & 1,2,3\\
 \gamma_5\gamma^\mu & \gamma_5\gamma_\mu & & 1,2,3\\
 \gamma_5\gamma^\mu D^\nu & \gamma_5\gamma_\nu D_\mu &\ho & 1,2,3\\
 \sigma^{\mu\nu} & \sigma_{\mu\nu} && 1,2,3\\
 \sigma^{\mu\nu}D^\rho & \sigma_{\mu\rho}D_\nu &\ho & 1,2,3\\
\bottomrule
\end{tabular}
\caption{Dirac operators \(\theta^i\) and \(\xi^i\) for contact terms of \(\mathcal{O}(q^0)\). An asterisk \ho\ in the column NR denotes structures which contribute at higher order in the non-relativistic expansion. The last column shows, to which flavor structures these operators contribute.} \label{tab:ct0}
\end{table}

\subsection[Contact terms of \texorpdfstring{$\mathcal{O}(q^1)$}{O(q\^{}1)}]{Contact terms of \boldmath{$\mathcal{O}(q^1)$}} \label{subsec:results-q1}

The Lagrangian of order \(\mathcal{O}(q^1)\) can be constructed by including a covariant derivative \(\hat D_2^\alpha\) or a field \(u^\alpha\), but not both.
When including the covariant derivative, one needs to consider again only terms of the type \(X^{1,2,3}\), due to the tracelessness of the baryon fields.
However, the restrictions through the equation of motion (see Appendix~\ref{sec:EOM}) and the special structure of the monomials \(X^{1}\) and \(X^{3}\) allow in the end for only two terms of type \(X^{2}\).
They read
\begin{align}
 \mathscr{L}^{(1)}_{BB} = \quad &b_1 \left(\hat D_2^\alpha\trace{\bar B_1\bar B_2(\gamma_5\gamma_\alpha D_\mu B_1)(\gamma_5\gamma^\mu B_2)} + \hat D_1^\alpha\trace{(\bar B_1\gamma_5\gamma^\mu) (\bar B_2 \Dl_\mu\gamma_5\gamma_\alpha) B_1B_2}\right) \notag \\
 +&b_2 \left(\hat D_2^\alpha\trace{\bar B_1\bar B_2(\gamma_5\gamma_\alpha D_{\mu\nu} B_1)(\gamma_5\gamma^\mu D^\nu B_2)} + \hat D_1^\alpha\trace{(\bar B_1\Dl_\nu\gamma_5\gamma^\mu) (\bar B_2 \Dl_{\mu\nu}\gamma_5\gamma_\alpha) B_1B_2}\right)\,,
\end{align}
with \(b_1\) and \(b_2\) new low-energy constants.
In the non-relativistic approximation these terms start to contribute at \(\mathcal{O}(q^2)\).
This behavior agrees with the fact, that parity conservation excludes any pure baryon-baryon contact terms at order \(\mathcal{O}(q^1)\).
The second term differs only from the first term, if more mesons are involved.

\bigskip
The other possibility to obtain terms at order \(\mathcal{O}(q^1)\) is to include the chiral building block \(u^\alpha\).
The corresponding terms can be of the type \(X^{1,2,3,4}\) with less than three flavor traces. They describe baryon-baryon contact interactions including additional Goldstone-boson fields. In total one can construct 67 terms for the Lagrangian,
\begin{equation} \begin{aligned}
 \mathscr{L}^{(1)}_{BB} = &\sum_{i=1}^{3} c_{1,i} \C^1_i + \sum_{i=1}^{7} c_{2,i} \C^2_i + \sum_{i=1}^{3} c_{3,i} \C^3_i + \sum_{i=1}^{7} c_{4,i} \C^4_i+ \sum_{i=1}^{7} c_{5,i} \C^5_i+ \sum_{i=1}^{7} c_{6,i} \C^6_i+ \sum_{i=1}^{7} c_{7,i} \C^7_i \\
 &+\sum_{i=1}^{3} c_{8,i} \C^8_i+ \sum_{i=1}^{3} c_{9,i} \C^9_i+ \sum_{i=1}^{3} c_{10,i} \C^{10}_i+ \sum_{i=1}^{3} c_{11,i} \C^{11}_i+ \sum_{i=1}^{7} c_{12,i} \C^{12}_i+ \sum_{i=1}^{7} c_{13,i} \C^{13}_i \,.
\end{aligned} \end{equation}
The general flavor structures for terms with one chiral building block \(A\) inserted are
\begin{align}
 \C^1_i &= \trace{\bar B_1 \extfield\theta^iB_1\bar B_2\xi^iB_2} + (-1)^\ci\trace{\bar B_1\thetal^i\extfield B_1\bar B_2\xil^iB_2} \label{eq:o2umu},\notag\\\notag
 \C^2_i &= \trace{\bar B_1 \theta^iB_1 \extfield\bar B_2\xi^iB_2} + (-1)^\ci\trace{\bar B_1\thetal^i B_1\bar B_2\xil^iB_2 \extfield},\\\notag
 \C^3_i &= \trace{\bar B_1 \extfield\xi^iB_1\bar B_2\theta^iB_2} + (-1)^\ci\trace{\bar B_1\xil^i\extfield B_1\bar B_2\thetal^iB_2}\displaybreak[0],\\[.3\baselineskip]\notag
 \C^4_i &= \trace{\bar B_1\extfield\bar B_2\theta^iB_1\xi^iB_2} + (-1)^\ci\trace{\bar B_1\xil^i\bar B_2\thetal^iB_1\extfield B_2},\\\notag
 \C^5_i &= \trace{\bar B_1\bar B_2\extfield\theta^iB_1\xi^iB_2} + (-1)^\ci\trace{\bar B_1\xil^i\bar B_2\thetal^i\extfield B_1 B_2},\\\notag
 \C^6_i &= \trace{\bar B_1\extfield\bar B_2\xi^iB_1\theta^iB_2} + (-1)^\ci\trace{\bar B_1\thetal^i\bar B_2\xil^iB_1\extfield B_2},\\\notag
 \C^7_i &= \trace{\bar B_1\bar B_2\theta^iB_1\xi^iB_2\extfield} + (-1)^\ci\trace{\bar B_1\xil^i\bar B_2\thetal^iB_1 B_2\extfield}\displaybreak[0],\\[.3\baselineskip]\notag
 \C^8_i &= \trace{\bar B_1\extfield\theta^iB_1}\trace{\bar B_2\xi^iB_2} + (-1)^\ci\trace{\bar B_1\thetal^i\extfield B_1}\trace{\bar B_2\xil^iB_2},\\\notag
 \C^9_i &= \trace{\bar B_1\theta^iB_1\extfield}\trace{\bar B_2\xi^iB_2} + (-1)^\ci\trace{\bar B_1\thetal^iB_1\extfield}\trace{\bar B_2\xil^iB_2},\\\notag
 \C^{10}_i &= \trace{\bar B_1\extfield\xi^iB_1}\trace{\bar B_2\theta^iB_2} + (-1)^\ci\trace{\bar B_1\xil^i\extfield B_1}\trace{\bar B_2\thetal^iB_2},\\\notag
 \C^{11}_i &= \trace{\bar B_1\xi^iB_1\extfield}\trace{\bar B_2\theta^iB_2} + (-1)^\ci\trace{\bar B_1\xil^iB_1\extfield}\trace{\bar B_2\thetal^iB_2}\displaybreak[0]\\[.3\baselineskip]\notag
 \C^{12}_i &= \trace{\bar B_1\extfield}\trace{\theta^iB_1\bar B_2\xi^iB_2} + (-1)^\ci\trace{\bar B_2\xil^iB_2\bar B_1\thetal^i}\trace{B_1\extfield} ,\\\notag
 \C^{13}_i &= \trace{\bar B_1\extfield}\trace{\xi^iB_1\bar B_2\theta^iB_2} + (-1)^\ci\trace{\bar B_2\thetal^iB_2\bar B_1\xil^i}\trace{B_1\extfield} \displaybreak[0],\\[.3\baselineskip]\notag
 \C^{14}_i &= \trace{\bar B_1 \theta^iB_1\bar B_2 \xi^iB_2}\trace{A} + (-1)^\ci\trace{\bar B_1\thetal^i B_1\bar B_2  \xil^iB_2}\trace{A} ,\\\notag
 \C^{15}_i &= \trace{\bar B_1\bar B_2\theta^iB_1\xi^iB_2}\trace{A} + (-1)^\ci\trace{\bar B_1\xil^i\bar B_2\thetal^iB_1 B_2}\trace{A} ,\\
 \C^{16}_i &= \trace{\bar B_1\theta^iB_1}\trace{\bar B_2\xi^iB_2}\trace{A} + (-1)^\ci\trace{\bar B_1\thetal^i B_1}\trace{\bar B_2\xil^iB_2}\trace{A} \,.
\end{align}
In the present case \(A=u^\alpha\), and the operators \(\theta^i\), \(\xi^i\) and the corresponding exponents \(c_i\) are given in Table~\ref{tab:ct1-2}.
Not all operators contribute to each flavor structure, e.g., if \(X^C=-X+\text{h.o.}\), or if one flavor structure is equal to another one up to higher order terms.
For the same reasons, the exchanged combinations with \(\theta^i\leftrightarrow\xi^i\) need not to be considered for all flavor structures.
The flavor structures \(\C^{14-16}_i\) do not appear, since \(u^\alpha\) is traceless.
All Dirac operators in Table~\ref{tab:ct1-2}, except the combinations \(\mathbbm1 \otimes \gamma_5\gamma_\alpha\) and \(\gamma_5\gamma^{\mu} \otimes {\sigma_\alpha}_{\mu}\), contribute in the non-relativistic expansion first at \(\mathcal{O}(q^2)\).
These criteria lead to 20 terms in the non-relativistic power counting.
It is worth to note that in the two-flavor case (with pions and nucleons only) one gets from this list of terms the much used \(4N\pi\) contact vertex proportional to the low-energy constant \(c_D\). It determines the mid-range \(1\pi\)-exchange component of the leading order chiral three-nucleon interaction \cite{Epelbaum2012}.

If more mesons are involved in addition to the Dirac operators in Table~\ref{tab:ct1-2} the same operators with one contracted pair of covariant derivatives have to be included, e.g.\ \(\mathbbm{1}\cdot {D_\alpha}^{\mu\nu} \otimes \gamma_5\gamma_\mu D_\nu\).
Their properties (\(\ci\), NR, \(\C^j\)) are the same as for the ones without the contracted pair in Table~\ref{tab:ct1-2}.

\begin{table}[htb]
\centering
\begin{tabular}{>{\(}c<{\)}>{\(}c<{\)}>{\(}c<{\)}>{\(}c<{\)}c}
\toprule
 \theta^i & \xi^i & \ci & \text{NR} & contributes to \(\C^j\) with \(A=u^\alpha\)\\
\midrule
 \mathbbm{1}\cdot {D_\alpha}^\mu & \gamma_5\gamma_\mu & 0 & \ho & 1-13\\
 \mathbbm{1}\cdot D^{\mu} & \gamma_5\gamma_\mu D_{\alpha} & 0 & \ho & 1-13\\
 \mathbbm{1} & \gamma_5\gamma_{\alpha} & 0 & & 1-13\\
 \mathrm i\ \gamma_5\gamma^{\mu}{D_\alpha}^\nu & \sigma_{\mu\nu} & 1 & \ho & 2,4,5,6,7,12,13\\
 \mathrm i\ \gamma_5\gamma^{\mu} & \sigma_{\alpha\mu} & 1 & & 2,4,5,6,7,12,13\\
 \mathrm i\ \gamma_5\gamma^{\mu}D^{\nu} & \sigma_{\alpha\nu} D_\mu & 1 & \ho & 2,4,5,6,7,12,13\\
 \mathrm i\ \gamma_5\gamma^{\mu}D^{\nu} & \sigma_{\mu\nu}D_\alpha & 1 & \ho & 2,4,5,6,7,12,13\\
\bottomrule
\end{tabular}
\caption{Dirac operators \(\theta^i\) and \(\xi^i\) for contact terms of \(\mathcal{O}(q^1)\) with one field \(u^\alpha\), and associated charge conjugation exponents \(c_i\). An asterisk \ho\ in the column NR denotes structures which are of higher order in the non-relativistic power counting. The last column shows, to which flavor structures these operators contribute.} \label{tab:ct1-2}
\end{table}

\subsection[Contact terms of \texorpdfstring{$\mathcal{O}(q^2)$}{O(q\^{}2)}]{Contact terms of \boldmath{$\mathcal{O}(q^2)$}} \label{subsec:results-q2}

In the following we construct the baryon-baryon contact Lagrangian at \(\mathcal{O}(q^2)\).
In the non-relativistic power counting the terms including Dirac operators marked by \ho\ do not contribute in a calculation up to \(\mathcal{O}(q^2)\). Nevertheless, we have decided to include these terms for the sake of completeness and in order to give a complete description of the contact terms in the covariant power counting.

\subsubsection*{Terms without external fields}

The first contributions to the Lagrangian of \(\mathcal{O}(q^2)\) comes from terms with two derivatives of baryon bilinears and no external fields. One obtains 18 such terms:
\begin{equation}
 \mathscr{L}^{(2)}_{BB} = \sum_{i=1}^{6} \left( d_{1,i} \D^1_i +  d_{2,i} \D^2_i +  d_{3,i} \D^3_i\right) \,,
\end{equation}
which are similar to the \(\mathcal{O}(q^0)\) terms,
\begin{equation} \begin{aligned}
 \D^1_i &= \hat D_2^{\alpha\beta}\trace{\bar B_1\theta^iB_1\bar B_2\xi^iB_2} + \hat D_2^{\alpha\beta}\trace{\bar B_1\thetal^iB_1\bar B_2\xil^iB_2},\\
 \D^2_i &= \hat D_2^{\alpha\beta}\trace{\bar B_1\bar B_2\theta^iB_1\xi^iB_2} + \hat D_1^{\alpha\beta}\trace{\bar B_1\xil^i\bar B_2\thetal^iB_1B_2},\\
 \D^3_i &= \hat D_2^{\alpha\beta}\trace{\bar B_1\theta^iB_1}\trace{\bar B_2\xi^iB_2} + \hat D_2^{\alpha\beta}\trace{\bar B_1\thetal^iB_1}\trace{\bar B_2\xil^iB_2} \,,
\end{aligned} \end{equation}
with the operators \(\theta^i\), \(\xi^i\) given in Table~\ref{tab:ct2-1}.
The structures \(g_{\alpha\beta} \gamma_5\gamma^\mu D^\nu \otimes \gamma_5\gamma_\nu D_\mu\) and \(g_{\alpha\beta} \sigma^{\mu\nu}D^\rho \otimes \sigma_{\mu\rho}D_\nu\) contribute in the non-relativistic counting at \(\mathcal{O}(q^3)\) or higher.
Therefore, one obtains at order \(\mathcal{O}(q^2)\) 12 relevant terms in the non-relativistic power counting.

\begin{table}[tb]
\centering
\begin{tabular}{>{\(}c<{\)}>{\(}c<{\)}>{\(}c<{\)}c}
\toprule
 \theta^i & \xi^i & \text{NR} & contributes to \(\D^j\)\\
\midrule
 g_{\alpha\beta} \mathbbm{1} & \mathbbm{1} & &1,2,3\\
 g_{\alpha\beta} \gamma_5\gamma^\mu & \gamma_5\gamma_\mu & &1,2,3\\
 g_{\alpha\beta} \gamma_5\gamma^\mu D^\nu & \gamma_5\gamma_\nu D_\mu & \ho & 1,2,3\\
 g_{\alpha\beta} \sigma^{\mu\nu} & \sigma_{\mu\nu} & & 1,2,3\\
 g_{\alpha\beta} \sigma^{\mu\nu}D^\rho & \sigma_{\mu\rho}D_\nu & \ho & 1,2,3\\
 \gamma_5\gamma_\alpha & \gamma_5\gamma_\beta & & 1,2,3\\
\bottomrule
\end{tabular}
\caption{Dirac operators \(\theta^i\) and \(\xi^i\) for contact terms of \(\mathcal{O}(q^2)\) without external fields. An asterisk \ho\ in the column NR indicates structures which are  at higher order in non-relativistic power counting. The last column shows, to which flavor structures these operators contribute.} \label{tab:ct2-1}
\end{table}

\subsubsection*{Terms including the external fields {\boldmath $\chi_\pm$}}

The terms including the external fields \(\chi_\pm\) are similar to the \(\mathcal{O}(q^1)\) terms including the field \(u_\mu\).
When setting the external scalar field equal to the quark mass matrix, these terms describe chiral symmetry breaking contact interactions. For \(\chi_+\) one finds in total 55 terms and for \(\chi_-\) one has in total 24 terms.
The Lagrangians for both cases read,
\begin{equation}
 \mathscr{L}^{(2)}_{BB} = \sum_{i,j} c^+_{j,i} \C^j_i(A\to\chi_+)\,,\qquad
 \mathscr{L}^{(2)}_{BB} = \sum_{i,j} c^-_{j,i} \C^j_i(A\to\chi_-)\,,
\end{equation}
with the flavor structures \(\C^j\) given in Eqs.~\eqref{eq:o2umu}.
The operators \(\theta^i\) and \(\xi^i\) for one insertion of \(\chi_+\) and \(\chi_-\) are given in Table~\ref{tab:ct2-2} and Table~\ref{tab:ct2-3}, respectively.
In the non-relativistic power counting the number of \(\chi_+\) terms reduces to 33 and all \(\chi_-\) terms are at least of order \(\mathcal{O}(q^3)\).

\begin{table}[htb]
\centering
\begin{tabular}{>{\(}c<{\)}>{\(}c<{\)}>{\(}c<{\)}>{\(}c<{\)}c}
\toprule
 \theta^i & \xi^i & \ci & \text{NR} & contributes to \(\C^j\)\ with \(A=\chi_+\)\\
\midrule
 \mathbbm{1} & \mathbbm{1} & 0 & & 1,2,4,5,7,8,9,12,14,15,16\\
 \gamma_5\gamma^\mu & \gamma_5\gamma_\mu & 0 & & 1,2,4,5,7,8,9,12,14,15,16\\
 \gamma_5\gamma^\mu D^\nu & \gamma_5\gamma_\nu D_\mu & 0 & \ho & 1,2,4,5,7,8,9,12,14,15,16\\
 \sigma^{\mu\nu} & \sigma_{\mu\nu} & 0 & & 1,2,4,5,7,8,9,12,14,15,16\\
 \sigma^{\mu\nu}D^\rho & \sigma_{\mu\rho}D_\nu & 0 & \ho & 1,2,4,5,7,8,9,12,14,15,16\\
\bottomrule
\end{tabular}
\caption{Dirac operators \(\theta^i\) and \(\xi^i\) for contact terms of \(\mathcal{O}(q^2)\) with \(\chi_+\). An asterisk \ho\ in the column NR indicates structures which are  at higher order in non-relativistic power counting. The last column shows, to which flavor structures these operators contribute.} \label{tab:ct2-2}
\end{table}
\begin{table}[htb]
\centering
\begin{tabular}{>{\(}c<{\)}>{\(}c<{\)}>{\(}c<{\)}>{\(}c<{\)}c}
\toprule
 \theta^i & \xi^i & \ci & \text{NR} & contributes to \(\C^j\)\ with \(A=\chi_-\)\\
\midrule
 \mathrm i\mathbbm{1}\cdot D^{\mu} & \gamma_5\gamma_\mu & 0 & \ho & 2,4,5,6,7,12,13,15\\
 \gamma_5\gamma^\mu D^\nu & \sigma_{\mu\nu} & 1 & \ho & 1-16\\
\bottomrule
\end{tabular}
\caption{Dirac structures \(\theta^i\) and \(\xi^i\) for contact terms of \(\mathcal{O}(q^2)\) with \(\chi_-\). An asterisk \ho\ in the column NR indicates structures which are at higher order in non-relativistic power counting. The last column shows, to which flavor structures these operators contribute.} \label{tab:ct2-3}
\end{table}

\subsubsection*{Terms including the fields {\boldmath $f^\pm_{\alpha\beta}\) and \(h_{\alpha\beta}$}}

When using the traceless chiral building blocks \(f_\pm^{\alpha\beta}\) and \(h^{\alpha\beta}\), which count of order \(\mathcal O(q^2)\), one obtains for each a contact Lagrangian,
\begin{equation}
 \mathscr{L}^{(2)}_{BB} = \sum_{i,j} c^\prime_{j,i} \C^j_i\,,
\end{equation}
with the (first thirteen) flavor structures \(\C^j,\ (j=1,\dots,13)\) listed in Eq.~\eqref{eq:o2umu} and the substitution \(A\to f_+^{\alpha\beta},f_-^{\alpha\beta},\,h^{\alpha\beta}\).
The Dirac operators \(\theta^i\) and \(\xi^i\) for \(A=f_+^{\alpha\beta}\) are given in Table~\ref{tab:ct2-p0}.
Column 4 in that table gives the additional factor \(\mathrm{i}\), if it is necessary for recovering hermiticity. Column 5 gives the corresponding charge conjugation exponent \(\ci\) and column 6 shows the flavor structures to which the occurring Dirac operators can contribute.
Table~\ref{tab:ct2-p1} gives the same information for the cases \(A=f_-^{\alpha\beta}\) and \(A=h^{\alpha\beta}\).
One obtains in total 127, 137 and 139 terms with one external field \(f_+^{\alpha\beta}\), \(f_-^{\alpha\beta}\) and \(h^{\alpha\beta}\), respectively.
Many of the Dirac operators in Table~\ref{tab:ct2-p0} and Table~\ref{tab:ct2-p1} contribute in the non-relativistic counting first at \(\mathcal{O}(q^3)\), and these are indicated by an asterisk \ho\ in the column NR. Examples for these are \(\gamma_5\gamma_\alpha D^\mu \otimes {\sigma_{\beta\mu}}\) and \({\epsilon_{\alpha\beta}}^{\delta\rho}\mathbbm1 D_\delta \otimes \mathbbm1 D_\rho\).
The number of \(\mathcal{O} (q^2)\) terms in non-relativistic counting reduces then to 33, 40 and 40 terms with one external field \(f_+^{\alpha\beta}\), \(f_-^{\alpha\beta}\) and \(h^{\alpha\beta}\), respectively.

\subsubsection*{Terms with one field {\boldmath $u_\mu$}}

The terms of order \(\mathcal{O}(q^2)\) with one field \(u^\alpha\) and hence with an additional covariant derivative \(\hat D_2^\beta\), have the same flavor structure as the \(\mathcal{O}(q^1)\) terms with one field \(u^\alpha\). The pertinent part of the contact Lagrangian is
\begin{equation}
 \mathscr{L}^{(2)}_{BB} = \sum_{i,j} e_{j,i} \E^j_i \,,
\end{equation}
with the flavor structures
\begin{align}
 \E^1_i &= \hat D_2^\beta\trace{\bar B_1 u^\alpha\theta^iB_1\bar B_2\xi^iB_2} + (-1)^\ci\hat D_2^\beta\trace{\bar B_1\thetal^iu^\alpha B_1\bar B_2\xil^iB_2}\notag,\\\notag\label{eq:o2E}
 \E^2_i &= \hat D_2^\beta\trace{\bar B_1 \theta^iB_1 u^\alpha\bar B_2\xi^iB_2} + (-1)^\ci\hat D_2^\beta\trace{\bar B_1\thetal^i B_1\bar B_2\xil^iB_2 u^\alpha},\\\notag
 \E^3_i &= \hat D_2^\beta\trace{\bar B_1 u^\alpha\xi^iB_1\bar B_2\theta^iB_2} + (-1)^\ci\hat D_2^\beta\trace{\bar B_1\xil^iu^\alpha B_1\bar B_2\thetal^iB_2}\displaybreak[0],\\[.3\baselineskip]\notag
 \E^4_i &= \hat D_2^\beta\trace{\bar B_1u^\alpha\bar B_2\theta^iB_1\xi^iB_2} + (-1)^\ci\hat D_1^\beta\trace{\bar B_1\xil^i\bar B_2\thetal^iB_1u^\alpha B_2},\\\notag
 \E^5_i &= \hat D_2^\beta\trace{\bar B_1\bar B_2u^\alpha\theta^iB_1\xi^iB_2} + (-1)^\ci\hat D_1^\beta\trace{\bar B_1\xil^i\bar B_2\thetal^iu^\alpha B_1 B_2},\\\notag
 \E^6_i &= \hat D_2^\beta\trace{\bar B_1u^\alpha\bar B_2\xi^iB_1\theta^iB_2} + (-1)^\ci\hat D_1^\beta\trace{\bar B_1\thetal^i\bar B_2\xil^iB_1u^\alpha B_2},\\\notag
 \E^7_i &= \hat D_2^\beta\trace{\bar B_1\bar B_2\theta^iB_1\xi^iB_2u^\alpha} + (-1)^\ci\hat D_1^\beta\trace{\bar B_1\xil^i\bar B_2\thetal^iB_1 B_2u^\alpha}\displaybreak[0],\\[.3\baselineskip]\notag
 \E^8_i &= \hat D_2^\beta\trace{\bar B_1u^\alpha\theta^iB_1}\trace{\bar B_2\xi^iB_2} + (-1)^\ci\hat D_2^\beta\trace{\bar B_1\thetal^iu^\alpha B_1}\trace{\bar B_2\xil^iB_2},\\\notag
 \E^9_i &= \hat D_2^\beta\trace{\bar B_1\theta^iB_1u^\alpha}\trace{\bar B_2\xi^iB_2} + (-1)^\ci\hat D_2^\beta\trace{\bar B_1\thetal^iB_1u^\alpha}\trace{\bar B_2\xil^iB_2},\\\notag
 \E^{10}_i &= \hat D_2^\beta\trace{\bar B_1u^\alpha\xi^iB_1}\trace{\bar B_2\theta^iB_2} + (-1)^\ci\hat D_2^\beta\trace{\bar B_1\xil^iu^\alpha B_1}\trace{\bar B_2\thetal^iB_2},\\\notag
 \E^{11}_i &= \hat D_2^\beta\trace{\bar B_1\xi^iB_1u^\alpha}\trace{\bar B_2\theta^iB_2} + (-1)^\ci\hat D_2^\beta\trace{\bar B_1\xil^iB_1u^\alpha}\trace{\bar B_2\thetal^iB_2}\displaybreak[0],\\[.3\baselineskip]\notag
 \E^{12}_i &= \hat D_2^\beta\trace{\bar B_1u^\alpha}\trace{\theta^iB_1\bar B_2\xi^iB_2} + (-1)^\ci\hat D_2^\beta\trace{\bar B_2\xil^iB_2\bar B_1\thetal^i}\trace{B_1u^\alpha}, \\
 \E^{13}_i &= \hat D_2^\beta\trace{\bar B_1u^\alpha}\trace{\xi^iB_1\bar B_2\theta^iB_2} + (-1)^\ci\hat D_2^\beta\trace{\bar B_2\thetal^iB_2\bar B_1\xil^i}\trace{B_1u^\alpha} \,.
\end{align}
The allowed Dirac operators \(\theta^i\) and \(\xi^i\) are given in columns 10-12 of Table~\ref{tab:ct2-p1}.
One obtains 82 such terms in covariant power counting out of which 14 remain in non-relativistic power counting.

\subsubsection*{Terms with two fields {\boldmath $u_\mu$} in the combination {\boldmath $[u_\alpha,u_\beta]_\pm$}}

When considering terms with two adjacent fields \(u_\alpha\) and \(u_\beta\) as the building block \(A\) to be inserted into the flavor structures \(\C^j_i\) (see Eq.~\eqref{eq:o2umu}), the two combinations \(A=\{u_\alpha,u_\beta\}\) and \(A=[u_\alpha,u_\beta]\) are to be considered separately.
The pertinent part of the contact Lagrangian is,
\begin{equation}
 \mathscr{L}^{(2)}_{BB} = \sum_{i,j} c^{\prime\prime}_{j,i} \C^j_i\,,
\end{equation}
with the allowed Dirac operators \(\theta^i\) and \(\xi^i\) given in Table~\ref{tab:ct2-p0}.
The columns 7-9 give the possibilities for the anticommutator \(A=\{u_\alpha,u_\beta\}\) and the columns 10-12 for the commutator \(A=[u_\alpha,u_\beta]\).
In total there are 303 terms for \(A=\{u_\alpha,u_\beta\}\) and 127 terms for \(A=[u_\alpha,u_\beta]\).
In the non-relativistic power counting the number of such terms reduces to 125 terms for \(A=\{u_\alpha,u_\beta\}\) and 33 terms for \(A=[u_\alpha,u_\beta]\).

\subsubsection*{Terms with two fields {\boldmath $u_\mu$} at non-neighboring positions}

The last possibility is to have two fields \(u^\alpha\) and \(u^\beta\) at non-neighboring positions in the flavor traces.
This allows for a large number of new flavor structures \(\F_i^j\) with up to three flavor traces. The pertinent part of the contact Lagrangian reads
\begin{equation}
 \mathscr{L}^{(2)}_{BB} = \sum_{i,j} f_{j,i} \F^j_i\,,
\end{equation}
with the flavor structures
\begin{align}
 \F^1_i &= \trace{\bar B_1 u^\alpha\theta^iB_1u^\beta\bar B_2\xi^iB_2} + (-1)^\ci\trace{\bar B_1\thetal^iu^\alpha B_1\bar B_2\xil^iB_2 u^\beta}, \qquad \F^{1^\prime}_i=\F^1_i|_{\theta^i\leftrightarrow\xi^i}, \notag\\\label{eq:o2F}
 \F^2_i &= \trace{\bar B_1 u^\alpha\theta^iB_1\bar B_2u^\beta\xi^iB_2} + (-1)^\ci\trace{\bar B_1\thetal^iu^\alpha B_1\bar B_2\xil^iu^\beta B_2}, \qquad \F^{2^\prime}_i=\F^2_i|_{\theta^i\leftrightarrow\xi^i}, \notag\\
 \F^3_i &= \trace{\bar B_1 \theta^iB_1u^\alpha\bar B_2u^\beta\xi^iB_2} + (-1)^\ci\trace{\bar B_1\thetal^i B_1\bar B_2\xil^iu^\beta B_2u^\alpha}, \qquad \F^{3^\prime}_i=\F^3_i|_{\theta^i\leftrightarrow\xi^i}, \notag\\
 \F^4_i &= \trace{\bar B_1 \theta^iB_1u^\alpha\bar B_2\xi^iB_2u^\beta} + (-1)^\ci\trace{\bar B_1\thetal^i B_1u^\beta\bar B_2\xil^i B_2u^\alpha}\displaybreak[0],\notag\\[.3\baselineskip]
\F^5_i &= \trace{\bar B_1u^\alpha\bar B_2u^\beta\theta^iB_1\xi^iB_2} + (-1)^\ci\trace{\bar B_1\xil^i\bar B_2\thetal^iu^\beta B_1u^\alpha B_2}, \qquad \F^{5^\prime}_i=\F^5_i|_{\theta^i\leftrightarrow\xi^i}, \notag\\
\F^6_i &= \trace{\bar B_1u^\alpha\bar B_2\theta^iB_1u^\beta\xi^iB_2} + (-1)^\ci\trace{\bar B_1\xil^i u^\beta\bar B_2\thetal^i B_1u^\alpha B_2}, \qquad \F^{6^\prime}_i=\F^6_i|_{\theta^i\leftrightarrow\xi^i}, \notag\\
\F^7_i &= \trace{\bar B_1u^\alpha\bar B_2\theta^iB_1\xi^iB_2u^\beta} + (-1)^\ci\trace{\bar B_1\xil^i\bar B_2\thetal^i B_1u^\alpha B_2 u^\beta}, \qquad \F^{7^\prime}_i=\F^7_i|_{\theta^i\leftrightarrow\xi^i}, \notag\\
\F^8_i &= \trace{\bar B_1\bar B_2u^\alpha\theta^iB_1u^\beta\xi^iB_2} + (-1)^\ci\trace{\bar B_1\xil^iu^\beta\bar B_2\thetal^i u^\alpha B_1 B_2}, \qquad \F^{8^\prime}_i=\F^8_i|_{\theta^i\leftrightarrow\xi^i}, \notag\\
\F^9_i &= \trace{\bar B_1\bar B_2u^\alpha\theta^iB_1\xi^iB_2u^\beta} + (-1)^\ci\trace{\bar B_1\xil^i\bar B_2\thetal^i u^\alpha B_1 B_2u^\beta}\displaybreak[0],\notag\\[.3\baselineskip]
\F^{10}_i &= \trace{\bar B_1u^\alpha\theta^iB_1u^\beta}\trace{\bar B_2\xi^iB_2} + (-1)^\ci\trace{\bar B_1\thetal^iu^\alpha B_1u^\beta}\trace{\bar B_2\xil^iB_2}, \qquad \F^{10^\prime}_i=\F^{10}_i|_{\theta^i\leftrightarrow\xi^i}, \notag\\
\F^{11}_i &= \trace{\bar B_1u^\alpha\theta^iB_1}\trace{\bar B_2u^\beta\xi^iB_2} + (-1)^\ci\trace{\bar B_1\thetal^iu^\alpha B_1}\trace{\bar B_2\xil^iu^\beta B_2}, \qquad \F^{11^\prime}_i=\F^{11}_i|_{\theta^i\leftrightarrow\xi^i}, \notag\\
\F^{12}_i &= \trace{\bar B_1u^\alpha\theta^iB_1}\trace{\bar B_2\xi^iB_2u^\beta} + (-1)^\ci\trace{\bar B_1\thetal^iu^\alpha B_1}\trace{\bar B_2\xil^i B_2 u^\beta}, \qquad \F^{12^\prime}_i=\F^{12}_i|_{\theta^i\leftrightarrow\xi^i}, \notag\\
\F^{13}_i &= \trace{\bar B_1\theta^iB_1u^\alpha}\trace{\bar B_2\xi^iB_2u^\beta} + (-1)^\ci\trace{\bar B_1\thetal^i B_1u^\alpha}\trace{\bar B_2\xil^i B_2 u^\beta}, \qquad \F^{13^\prime}_i=\F^{13}_i|_{\theta^i\leftrightarrow\xi^i}\displaybreak[0],\notag\\[.3\baselineskip]
\F^{14}_i &= \trace{\bar B_1u^\alpha}\trace{\theta^iB_1u^\beta\bar B_2\xi^iB_2} + (-1)^\ci\trace{\bar B_2\xil^i B_2u^\beta\bar B_1\thetal^i}\trace{B_1u^\alpha}, \qquad \F^{14^\prime}_i=\F^{14}_i|_{\theta^i\leftrightarrow\xi^i}, \notag\\
\F^{15}_i &= \trace{\bar B_1u^\alpha}\trace{\theta^iB_1\bar B_2u^\beta\xi^iB_2} + (-1)^\ci\trace{\bar B_2\xil^i u^\beta B_2\bar B_1\thetal^i}\trace{B_1u^\alpha}, \qquad \F^{15^\prime}_i=\F^{15}_i|_{\theta^i\leftrightarrow\xi^i}, \notag\\
\F^{16}_i &= \trace{\bar B_1u^\alpha}\trace{\theta^iB_1\bar B_2\xi^iB_2u^\beta} + (-1)^\ci\trace{\bar B_2\xil^i B_2\bar B_1\thetal^iu^\beta}\trace{B_1u^\alpha}, \qquad \F^{16^\prime}_i=\F^{16}_i|_{\theta^i\leftrightarrow\xi^i}\displaybreak[0],\notag\\[.3\baselineskip]
\F^{17}_i &= \trace{\bar B_1\xi^iB_1}\trace{\bar B_2u^\alpha}\trace{\theta^iB_2u^\beta} + (-1)^\ci\trace{\bar B_1\xil^i B_1}\trace{\bar B_2\thetal^iu^\beta}\trace{B_2u^\alpha}, \qquad \F^{17^\prime}_i=\F^{17}_i|_{\theta^i\leftrightarrow\xi^i}\displaybreak[0],\notag\\[.3\baselineskip]
\F^{18}_i &= \trace{\bar B_1u^\alpha}\trace{\bar B_2u^\beta}\trace{\theta^i B_1\xi^i B_2} + (-1)^\ci\trace{\bar B_2 \thetal^i \bar B_1 \xil^i}\trace{B_2 u^\beta}\trace{B_1 u^\alpha}, \qquad \F^{18^\prime}_i=\F^{18}_i|_{\theta^i\leftrightarrow\xi^i} \,.
\end{align}
The Dirac operators \(\theta^i\) and \(\xi^i\) are given in columns 13-15 of Table~\ref{tab:ct2-p0}.
One counts in total 817 such terms.
In the non-relativistic power counting their number reduces to 276.
A further reduction of the flavor structures might be possible when applying the Lagrangian to definite processes.

\newcommand{\Call}{\text{1-13}}
\newcommand{\Cnonis}{\parbox{10ex}{\centering\smallskip 1,2,4,5,\newline 7,8,9,12\smallskip}}
\newcommand{\Cnostar}{\parbox{10ex}{\centering\smallskip 2,4,5,6,\newline 7,12,13\smallskip}}
\newcommand{\Cnostarnis}{\parbox{10ex}{\centering\smallskip 2,4,5,\newline 7,12\smallskip}}
\newcommand{\Cnostarif}{\parbox{12ex}{\centering\smallskip 1,4,8,\newline 9,12\smallskip}}
\newcommand{\Callfull}{\text{1-16}}
\newcommand{\Cnostarfull}{\parbox{10ex}{\centering\smallskip 2,4,5,6,7,\newline 12,13,15\smallskip}}
\newcommand{\Cnonisfull}{\parbox{15ex}{\centering\smallskip 1,2,4,5,7,8,\newline 9,12,14,15,16\smallskip}}
\newcommand{\Eall}{\text{1-13}}
\newcommand{\Enostar}{\parbox{10ex}{\centering\smallskip 2,4,5,6,\newline 7,12,13\smallskip}}
\newcommand{\Enostarif}{\parbox{10ex}{\centering\smallskip 1,4,5,7,\newline 8,9,12 \smallskip}}
\newcommand{\Fsabe}{1\text-18,7',10',12',14'\text-16',17'}
\newcommand{\Fsabo}{\parbox{20ex}{\centering\smallskip$1,3,5\text-9,14\text-16,18,$\newline $7',14'\text-16'$\smallskip}}
\newcommand{\Faabe}{\parbox{20ex}{\centering\smallskip$1\text-3,5\text-16,18,$\newline $7',10',12',14'\text-16'$\smallskip}}
\newcommand{\Faabo}{\parbox{20ex}{\centering\smallskip$1,3\text-9,14\text-18,$\newline $7',14'\text-17'$\smallskip}}
\newcommand{\Fevche}{\parbox{20ex}{\centering\smallskip$1,2,4\text-7,9\text-18,$\newline $1',2',5'\text-7',10'\text-18'$\smallskip}}
\newcommand{\Fevcho}{\parbox{20ex}{\centering\smallskip$1,4,5,7,9,14\text-18,$\newline $1',5',7',14'\text-18'$\smallskip}}
\newcommand{\Fche}{1\text-18}
\newcommand{\Fexabe}{\parbox{20ex}{\centering\smallskip$7,9,10,12,14\text-17,$\newline $7',10',12',14'\text-17'$\smallskip}}
\newcommand{\Fexabo}{\parbox{20ex}{\centering\smallskip$7,9,14\text-17,$\newline $7',14'\text-17'$\smallskip}}
\newcommand{\Fnoe}{1\text-18,1'\text-3',5'\text-8',10'\text-18'}
\newcommand{\Fnoo}{\parbox{20ex}{\centering\smallskip$1,3\text-9,14\text-18,$\newline $1',3',5'\text-8',14'\text-18'$\smallskip}}
\newcommand{\nocontr}{& & &}
\newcommand{\noprefactor}{1}

\begin{table}[p]
\setlength{\tabcolsep}{3pt}
\centering\footnotesize
\resizebox*{!}{.925\textheight}{
\begin{tabular}{>{\(}c<{\)}>{\(}c<{\)}>{\(}c<{\)}>{\(}c<{\)}>{\(}c<{\)}>{\(}c<{\)}>{\(}c<{\)}>{\(}c<{\)}>{\(}c<{\)}>{\(}c<{\)}>{\(}c<{\)}>{\(}c<{\)}>{\(}c<{\)}>{\(}c<{\)}>{\(}c<{\)}}
\toprule
 & & & \multicolumn{3}{c}{\(\C^j\) with \(A=f_+^{\alpha\beta}\)} & \multicolumn{3}{c}{\(\C^j\) with \(A=\{u^\alpha,u^\beta\}\)} & \multicolumn{3}{c}{\(\C^j\) with \(A=[u^\alpha,u^\beta]\)} & \multicolumn{3}{c}{\(\F^j\)} \\
 \theta^i & \xi^i & \text{NR} & \text{factor} & \ci & \C^j & \text{factor} & \ci & \C^j & \text{factor} & \ci & \C^j & \text{factor} & \ci & \F^j \\
\cmidrule(r){1-3}\cmidrule(lr){4-6}\cmidrule(lr){7-9}\cmidrule(lr){10-12}\cmidrule(l){13-15}
 g_{\alpha\beta} \mathbbm{1} & \mathbbm{1}&    \nocontr  & \noprefactor & 0 & \Cnonisfull   \nocontr   & \noprefactor & 0 & \Fche \\
 \mathbbm{1} D_{\alpha\beta} & \mathbbm{1}&    \nocontr  & \noprefactor & 0 & \Callfull   \nocontr   & \noprefactor & 0 & \Fsabe \\
 \mathbbm{1} D_\alpha & \mathbbm{1}D_\beta&  &    \mathrm i & 1 & \Cnostarnis     & \noprefactor & 0 & \Cnonisfull    & \noprefactor & 1 & \Cnostarnis    & \noprefactor & 0 & \Fevche \\
\cmidrule(r){1-3}\cmidrule(lr){4-6}\cmidrule(lr){7-9}\cmidrule(lr){10-12}\cmidrule(l){13-15}
 \mathbbm{1}{D_\alpha}^\mu & \sigma_{\beta\mu}&\ho    & \noprefactor & 0 & \Call   & \mathrm i & 1 & \Cnostarfull   & \mathrm i & 0 & \Call   & \mathrm i & 1 & \Fnoo \\
 \mathbbm{1} & \sigma_{\alpha\beta}&    & \noprefactor & 0 & \Call   \nocontr  & \mathrm i & 0 & \Call   & \mathrm i & 1 & \Faabo \\
 \mathbbm{1}D^\mu & \sigma_{\alpha\mu} D_\beta&\ho    & \noprefactor & 0 & \Call   & \mathrm i & 1 & \Cnostarfull   & \mathrm i & 0 & \Call   & \mathrm i & 1 & \Fnoo \\
 \mathbbm{1}{D_{\beta}}^\mu & \sigma_{\alpha\mu}&\ho    \nocontr   \nocontr   \nocontr   & \mathrm i & 1 & \Fexabo \\
 \mathbbm{1}D^\mu & \sigma_{\beta\mu} D_\alpha&\ho    \nocontr   \nocontr   \nocontr   & \mathrm i & 1 & \Fexabo \\
\cmidrule(r){1-3}\cmidrule(lr){4-6}\cmidrule(lr){7-9}\cmidrule(lr){10-12}\cmidrule(l){13-15}
 g_{\alpha\beta} \gamma_5\gamma^\mu & \gamma_5\gamma_\mu&    \nocontr   & \noprefactor & 0 & \Cnonisfull   \nocontr   & \noprefactor & 0 & \Fche \\
 g_{\alpha\beta} \gamma_5\gamma^\mu D^{\nu} & \gamma_5\gamma_\nu D_\mu&\ho    \nocontr   & \noprefactor & 0 & \Cnonisfull   \nocontr   & \noprefactor & 0 & \Fche \\
 \gamma_5\gamma_\alpha {D_\beta}^\mu & \gamma_5\gamma_\mu&\ho    & \mathrm i & 1 & \Cnostar   & \noprefactor & 0 & \Callfull   & \noprefactor & 1 & \Cnostar   & \noprefactor & 0 & \Fnoe \\
 \gamma_5\gamma_\alpha & \gamma_5\gamma_\beta&    & \mathrm i & 1 & \Cnostarnis   & \noprefactor & 0 & \Cnonisfull   & \noprefactor & 1 & \Cnostarnis   & \noprefactor & 0 & \Fevche \\
 \gamma_5\gamma_\alpha D^{\mu} & \gamma_5\gamma_\mu D_\beta&\ho    & \mathrm i & 1 & \Cnostar   & \noprefactor & 0 & \Callfull   & \noprefactor & 1 & \Cnostar   & \noprefactor & 0 & \Fnoe \\
 \gamma_5\gamma^\mu D_{\alpha\beta} & \gamma_5\gamma_\mu&    \nocontr   & \noprefactor & 0 & \Callfull   \nocontr   & \noprefactor & 0 & \Fsabe \\
 \gamma_5\gamma^\mu {D_{\alpha\beta}}^{\nu} & \gamma_5\gamma_\nu D_\mu&\ho    \nocontr   & \noprefactor & 0 & \Callfull   \nocontr   & \noprefactor & 0 & \Fsabe \\
 \gamma_5\gamma^\mu D_\alpha & \gamma_5\gamma_\mu D_\beta&    & \mathrm i & 1 & \Cnostarnis   & \noprefactor & 0 & \Cnonisfull   & \noprefactor & 1 & \Cnostarnis   & \noprefactor & 0 & \Fevche \\
 \gamma_5\gamma^\mu {D_\alpha}^{\nu} & \gamma_5\gamma_\nu {D}_{\beta\mu}&\ho    & \mathrm i & 1 & \Cnostarnis   & \noprefactor & 0 & \Cnonisfull   & \noprefactor & 1 & \Cnostarnis   & \noprefactor & 0 & \Fevche \\
 \gamma_5\gamma^\mu D_{\alpha} & \gamma_5\gamma_\beta D_\mu&\ho     \nocontr   \nocontr   \nocontr   & \noprefactor & 0 & \Fnoe \\
 \gamma_5\gamma_\beta {D_\alpha}^{\mu} & \gamma_5\gamma_\mu&\ho    \nocontr   \nocontr   \nocontr   & \noprefactor & 0 & \Fexabe \\
\cmidrule(r){1-3}\cmidrule(lr){4-6}\cmidrule(lr){7-9}\cmidrule(lr){10-12}\cmidrule(l){13-15}
 g_{\alpha\beta}\sigma^{\mu\nu} & \sigma_{\mu\nu}&    \nocontr   & \noprefactor & 0 & \Cnonisfull   \nocontr   & \noprefactor & 0 & \Fche \\
 g_{\alpha\beta}\sigma^{\mu\nu}D^{\rho} & \sigma_{\mu\rho} D_\nu&\ho    \nocontr  & \noprefactor & 0 & \Cnonisfull   \nocontr   & \noprefactor & 0 & \Fche \\
 {\sigma_\alpha}^{\mu}{D_\beta}^{\nu} & \sigma_{\mu\nu}&\ho    & \mathrm i & 1 & \Cnostar   & \noprefactor & 0 & \Callfull   & \noprefactor & 1 & \Cnostar   & \noprefactor & 0 & \Fnoe \\
 {\sigma_\alpha}^{\mu} & {\sigma}_{\beta\mu}&    & \mathrm i & 1 & \Cnostarnis   & \noprefactor & 0 & \Cnonisfull   & \noprefactor & 1 & \Cnostarnis   & \noprefactor & 0 & \Fevche \\
 {\sigma_\alpha}^{\mu}D^{\nu} & {\sigma}_{\beta\nu}D_\mu&\ho    & \mathrm i & 1 & \Cnostarnis   & \noprefactor & 0 & \Cnonisfull   & \noprefactor & 1 & \Cnostarnis   & \noprefactor & 0 & \Fevche \\
 {\sigma_\alpha}^{\mu}D^{\nu} & \sigma_{\mu\nu} D_\beta&\ho    & \mathrm i & 1 & \Cnostar   & \noprefactor & 0 & \Callfull   & \noprefactor & 1 & \Cnostar   & \noprefactor & 0 & \Fnoe \\
 \sigma^{\mu\nu}D_{\alpha\beta} & \sigma_{\mu\nu}&    \nocontr   & \noprefactor & 0 & \Callfull   \nocontr   & \noprefactor & 0 & \Fsabe \\
 \sigma^{\mu\nu}{D^{\rho}}_{\alpha\beta} & \sigma_{\mu\rho} D_\nu&\ho    \nocontr  & \noprefactor & 0 & \Callfull   \nocontr   & \noprefactor & 0 & \Fsabe \\
 \sigma^{\mu\nu}D_{\alpha} & \sigma_{\beta\mu} D_\nu&    \nocontr   \nocontr   \nocontr   & \noprefactor & 0 & \Fnoe \\
 \sigma^{\mu\nu}D_{\alpha} & \sigma_{\mu\nu} D_\beta&\ho    & \mathrm i & 1 & \Cnostarnis   & \noprefactor & 0 & \Cnonisfull   & \noprefactor & 1 & \Cnostarnis   & \noprefactor & 0 & \Fevche \\
 \sigma^{\mu\nu}{D^{\rho}}_\alpha & \sigma_{\mu\rho} D_{\nu\beta}&\ho    & \mathrm i & 1 & \Cnostarnis   & \noprefactor & 0 & \Cnonisfull   & \noprefactor & 1 & \Cnostarnis   & \noprefactor & 0 & \Fevche \\
 {\sigma_\beta}^{\mu}{D_\alpha}^{\nu} & \sigma_{\mu\nu}&\ho    \nocontr   \nocontr   \nocontr   & \noprefactor & 0 & \Fexabe \\
\cmidrule(r){1-3}\cmidrule(lr){4-6}\cmidrule(lr){7-9}\cmidrule(lr){10-12}\cmidrule(l){13-15}
 {\epsilon_{\alpha\beta}}^{\delta\rho}\mathbbm{1}{D_\delta}^\mu & \gamma_5\gamma_\mu D_\rho&\ho    & \mathrm i & 1 & \Call   \nocontr   & \noprefactor & 1 & \Call   & \noprefactor & 0 & \Faabo \\
 {\epsilon_{\alpha\beta}}^{\delta\rho}\gamma_5\gamma^\mu {D_\delta}^\nu & \sigma_{\mu\nu} D_\rho&\ho    & \noprefactor & 0 & \Cnostar   \nocontr   & \mathrm i & 0 & \Cnostar   & \mathrm i & 1 & \Faabe \\
\bottomrule
\end{tabular}
}
\caption{Dirac operators \(\theta^i\) and \(\xi^i\) for contact terms of \(\mathcal{O}(q^2)\) with even-parity external fields carrying two Lorentz indices.
The sets of flavor structures \(\C^j\) and \(\F^j\) are listed in Eqs.~\eqref{eq:o2umu} and \eqref{eq:o2F}.
Combinations that contribute in the non-relativistic power counting at \(\mathcal{O}(q^3)\) or higher are marked by an asterisk~\ho.
The columns ``factor'' indicate the cases where a prefactor \(\mathrm i\) is necessary for Hermitian conjugation invariance.
The columns ``\(c_i\)'' give the charge conjugation exponent \(\ci\).
The allowed flavor structures \(\C^j,\F^j\) are given by listing the indices \(j\) of the corresponding subset.
} \label{tab:ct2-p0}
\setlength{\tabcolsep}{6pt}
\end{table}

\begin{table}[htb!]
\setlength{\tabcolsep}{3pt}
\centering\footnotesize
\begin{tabular}{>{\(}c<{\)}>{\(}c<{\)}>{\(}c<{\)}>{\(}c<{\)}>{\(}c<{\)}>{\(}c<{\)}>{\(}c<{\)}>{\(}c<{\)}>{\(}c<{\)}>{\(}c<{\)}>{\(}c<{\)}>{\(}c<{\)}}
\toprule
 & & & \multicolumn{3}{c}{\(\C^j\) with \(A=f_-^{\alpha\beta}\)} & \multicolumn{3}{c}{\(\C^j\) with \(A=h^{\alpha\beta}\)} & \multicolumn{3}{c}{\(\E^j\)} \\
\theta^i & \xi^i & \text{NR} & \text{factor} & \ci & \C^j & \text{factor} & \ci & \C^j & \text{factor} & \ci & \E^j \\
\cmidrule(r){1-3}\cmidrule(lr){4-6}\cmidrule(lr){7-9}\cmidrule(lr){10-12}
 g_{\alpha\beta} \mathbbm{1} D^\mu & \gamma_5\gamma_\mu&\ho    \nocontr   \nocontr   & \noprefactor & 0 & \Enostar \\
 \mathbbm{1} {D_{\alpha\beta}}^{\mu} & \gamma_5\gamma_\mu&\ho    \nocontr   & \noprefactor & 0 & \Cnostar   \nocontr \\
 \mathbbm{1} D_{\alpha} & \gamma_5\gamma_\beta&    & \noprefactor & 0 & \Cnostar   & \noprefactor & 0 & \Cnostar   & \noprefactor & 0 & \Enostar \\
 \mathbbm{1} {D_\alpha}^{\mu} & \gamma_5\gamma_\mu D_\beta&\ho    & \noprefactor & 0 & \Cnostar   & \noprefactor & 0 & \Cnostar   \nocontr \\
 \mathbbm{1} & \gamma_5\gamma_\alpha D_\beta&    & \noprefactor & 0 & \Cnostar   & \noprefactor & 0 & \Cnostar   \nocontr \\
 \mathbbm{1}D^\mu & \gamma_5\gamma_\mu D_{\alpha\beta}&\ho    \nocontr   & \noprefactor & 0 & \Cnostar   \nocontr \\
 \mathbbm{1} & \gamma_5\gamma_\beta D_\alpha&    \nocontr   \nocontr   & \noprefactor & 0 & \Enostar \\
\cmidrule(r){1-3}\cmidrule(lr){4-6}\cmidrule(lr){7-9}\cmidrule(lr){10-12}
 g_{\alpha\beta}\gamma_5\gamma^\mu D^{\nu} & \sigma_{\mu\nu}&\ho    \nocontr   \nocontr   & \mathrm i & 1 & \Eall\\
 \gamma_5\gamma_\alpha D^\mu & {\sigma}_{\beta\mu}&\ho    & \mathrm i & 1 & \Call   & \mathrm i & 1 & \Call   \nocontr \\
 \gamma_5\gamma^\mu {D_{\alpha\beta}}^{\nu} & \sigma_{\mu\nu}&\ho    \nocontr   & \mathrm i & 1 & \Call   \nocontr \\
 \gamma_5\gamma^\mu D_{\alpha} & {\sigma}_{\beta\mu}&    & \mathrm i & 1 & \Call   & \mathrm i & 1 & \Call   \nocontr \\
 \gamma_5\gamma^\mu D^{\alpha\nu} & {\sigma^\beta}_\nu D_\mu&\ho    & \mathrm i & 1 & \Call   & \mathrm i & 1 & \Call   \nocontr \\
 \gamma_5\gamma^\mu {D_\alpha}^{\nu} & \sigma_{\mu\nu} D_\beta&\ho    & \mathrm i & 1 & \Call   & \mathrm i & 1 & \Call   \nocontr \\
 \gamma_5\gamma^\mu & \sigma_{\alpha\beta} D_\mu&\ho    & \mathrm i & 1 & \Call   \nocontr   \nocontr \\
 \gamma_5\gamma^\mu & {\sigma}_{\alpha\mu} D_\beta&    & \mathrm i & 1 & \Call   & \mathrm i & 1 & \Call   \nocontr \\
 \gamma_5\gamma^\mu D^\nu & {\sigma}_{\alpha\nu} {D}_{\beta\mu}&\ho    & \mathrm i & 1 & \Call   & \mathrm i & 1 & \Call   \nocontr \\
 \gamma_5\gamma^\mu D^\nu & \sigma_{\mu\nu} D_{\alpha\beta}&\ho    \nocontr   & \mathrm i & 1 & \Call   \nocontr \\
 \gamma_5\gamma_\beta D^\mu & {\sigma}_{\alpha\mu}&\ho    \nocontr   \nocontr   & \mathrm i & 1 & \Eall \\
\cmidrule(r){1-3}\cmidrule(lr){4-6}\cmidrule(lr){7-9}\cmidrule(lr){10-12}
 {\epsilon_{\alpha\beta}}^{\delta\rho}\mathbbm1 D_\delta & \mathbbm1 D_\rho&\ho    & \noprefactor & 0 & \Cnostarif   \nocontr   & \noprefactor & 0 & \Enostarif \\
 {\epsilon_{\alpha\beta}}^{\delta\rho}\gamma_5\gamma^\mu D_\delta & \gamma_5\gamma_\mu D_\rho&\ho    & \noprefactor & 0 & \Cnostarif   \nocontr   & \noprefactor & 0 & \Enostarif \\
 {\epsilon_{\alpha\beta}}^{\delta\rho}\gamma_5\gamma^\mu {D^{\nu}}_\delta & \gamma_5\gamma_\nu D_{\mu\rho}&\ho    & \noprefactor & 0 & \Cnostarif   \nocontr   & \noprefactor & 0 & \Enostarif \\
 {\epsilon_{\alpha\beta}}^{\delta\rho}\sigma^{\mu\nu} D_\delta & \sigma_{\mu\nu} D_\rho&\ho    & \noprefactor & 0 & \Cnostarif   \nocontr   & \noprefactor & 0 & \Enostarif \\
 {\epsilon_{\alpha\beta}}^{\delta\rho}\sigma^{\mu\nu} {D^\xi}_\delta & \sigma_{\mu\xi} {D_\nu}_\rho&\ho    & \noprefactor & 0 & \Cnostarif   \nocontr   & \noprefactor & 0 & \Enostarif \\
\bottomrule
\end{tabular}
\caption{Dirac operators \(\theta^i\) and \(\xi^i\) for contact terms of \(\mathcal{O}(q^2)\) with odd-parity external fields carrying two Lorentz indices.
The sets of flavor structures \(\C^j\) and \(\E^j\) are listed in Eqs.~\eqref{eq:o2umu} and \eqref{eq:o2E}.
Combinations that contribute in non-relativistic power counting at \(\mathcal{O}(q^3)\) or higher are marked by an asterisk~\ho.
The columns ``factor'' indicate the cases where a prefactor \(\mathrm i\) is necessary for Hermitian conjugation invariance.
The columns ``\(c_i\)'' give the charge conjugation exponent \(\ci\).
The allowed flavor structures \(\C^j,\E^j\) are given by listing the indices \(j\) of the corresponding subset.
} \label{tab:ct2-p1}
\setlength{\tabcolsep}{6pt}
\end{table}

\section{Application to baryon-baryon interactions} \label{sec:application}

In the case of contact terms relevant for the pure baryon-baryon scattering potentials, where no pseudoscalar mesons are involved, almost all external fields can be dropped.
All covariant derivatives \(D_\mu\) reduce to ordinary derivatives \(\partial_\mu\).
The only surviving external field is \(\chi_+\), which is then responsible for the inclusion of quark masses into the chiral Lagrangian:
\begin{equation}
 \frac{\chi_+}2 = \chi = 2B_0 \begin{pmatrix} m_u & 0 & 0 \\ 0 & m_d & 0 \\ 0 & 0 & m_s \end{pmatrix} \approx \begin{pmatrix} m_\pi^2 & 0 & 0 \\ 0 & m_\pi^2 & 0 \\ 0 & 0 & 2m_K^2-m_\pi^2 \end{pmatrix} \,,
\end{equation}
where in the last step the Gell-Mann--Oakes--Renner relation was used.
The corresponding terms provide the explicit SU(3) symmetry breaking contact potentials linear in the quark masses.
\enlargethispage{\baselineskip}
The subset of contact terms proportional to \(\trace{\chi_\pm}\) can be absorbed in the leading order terms, since the corresponding low-energy constants merely get shifted by a correction linear in the sum of the quark masses.

Using the results of Sec.~\ref{sec:results} one obtains the relevant linearly independent Lagrangians displayed in Table~\ref{tab:BBLagrangians},\footnote{In order to obtain a minimal set of Lagrangian terms, we have decomposed the emerging potentials into partial waves.
For each partial wave one gets a matrix which connects the Lagrangian constants with the different baryon-baryon channels. Lagrangian terms are considered as redundant if their omission does not reduce the rank of this matrix. In the case of the SU(3)-breaking terms we have done this reduction together with the leading SU(3) symmetric terms.}
which contribute in the non-relativistic power counting up to \(\mathcal O (q^2)\).
The first 28 terms contain only baryon fields and derivatives, and are therefore SU(3) symmetric.
The other 12 terms include the diagonal matrix \(\chi\) and produce explicit SU(3) symmetry breaking.
Again, the operator \(\hat \partial_i\) means, that the derivative acts only on baryon fields in the baryon bilinear \(i\).

After a non-relativistic expansion to \(\mathcal O (q^2)\) these terms lead to potentials in spin space.
We use the basis and partial wave decomposition formulas of Ref.~\cite{Polinder2006}:
\begin{align}
 P_1 &= 1\,,& \quad P_2 &= \vec\sigma_1\cdot\vec\sigma_2\,,\notag\displaybreak[0]\\
 P_3 &= (\vec\sigma_1\cdot\vec k\,)(\vec\sigma_2\cdot\vec k\,) - \frac13(\vec\sigma_1\cdot\vec\sigma_2)\vec k^{\,2} \,,& \quad P_4 &= \frac{\mathrm i}2(\vec\sigma_1+\vec\sigma_2)\cdot\vec n \,,\notag\displaybreak[0]\\
 P_5 &= (\vec\sigma_1\cdot\vec n\,)(\vec\sigma_2\cdot\vec n\,)\,,& \quad P_6 &= \frac{\mathrm i}2(\vec\sigma_1-\vec\sigma_2)\cdot\vec n \,,\notag\displaybreak[0]\\
 P_7 &= (\vec\sigma_1\cdot\vec q\,)(\vec\sigma_2\cdot\vec k\,) + (\vec\sigma_1\cdot\vec k\,)(\vec\sigma_2\cdot\vec q\,)\,,& \quad
 P_8 &= (\vec\sigma_1\cdot\vec q\,)(\vec\sigma_2\cdot\vec k\,) - (\vec\sigma_1\cdot\vec k\,)(\vec\sigma_2\cdot\vec q\,)\,,
\end{align}
with
\begin{equation}
 \vec q = \frac12(\vec p_f+\vec p_i)\,,\quad \vec k = \vec p_f - \vec p_i\,,\quad \vec n = \vec p_i\times\vec p_f\,.
\end{equation}
The momenta \(\vec p_f\) and \(\vec p_i\) are the initial and final state momenta in the center-of-mass frame.
For convenience we define additionally the combination \(P_9 = \vec\sigma_1\cdot \vec q\,\vec\sigma_2\cdot \vec q = (\frac{n^2}{k^2}-\frac13q^2)P_2 -\frac{q^2}{k^2}P_3-\frac1{k^2}P_5+\frac{p_f^2-p_i^2}{2k^2}P_7\).
In Table~\ref{tab:BBLagrangians} we show to which of these structures the terms in the Lagrangian contribute.
Note, that these are only the direct contributions.
Additional structures are obtained from contributions with exchanged final state baryons, where the spin exchange operator \(P_\sigma=\frac12\left(1+\vec\sigma_1\cdot\vec\sigma_2\right)\) is applied.

The basis elements \(P_6\) and \(P_8\) lead to spin singlet-triplet transitions (\({}^1P_1\leftrightarrow{}^3P_1\)). This comes only from the term~28 in Table~\ref{tab:BBLagrangians}, which gives rise to the antisymmetric spin-orbit operator \(P_6\) and via spin exchange to \(P_8 = 2\left. P_\sigma P_6\right|_{\vec p_f\rightarrow-\vec p_f}\). Therefore only one low-energy constant for the singlet-triplet mixing is present.
For the \(NN\) interaction such transitions are only possible when isospin breaking is included. Here even in the SU(3) limit, for some baryon channels this transition is possible.

In Table~\ref{tab:PWDBB} we present the non-vanishing transitions projected to the partial wave basis.
We recover the SU(3) relations of \cite{Polinder2006,Haidenbauer2013a,Polinder2007,Haidenbauer2010a}.
The constants are already redefined so that one sees the flavor content regarding the irreducible SU(3) representations
\( \mathbf8\otimes\mathbf8 = \mathbf{27}_s\oplus\mathbf{10}_a\oplus\mathbf{10^*}_a\oplus\mathbf{8}_s\oplus\mathbf{8}_a\oplus\mathbf1_s \).
The symmetric flavor representations \(\mathbf{27}_s,\ \mathbf{8}_s,\ \mathbf 1_s\) combine with the space-spin antisymmetric partial waves \({}^1S_0,\ {}^3P_0,\ {}^3P_1,\ {}^3P_2 \) with the 15 constants \(\tilde c_{{}^1S_0}^{27,8s,1}\), \(c_{{}^1S_0}^{27,8s,1}\), \(c_{{}^3P_0}^{27,8s,1}\), \(c_{{}^3P_1}^{27,8s,1}\) and \(c_{{}^3P_2}^{27,8s,1}\).
The antisymmetric flavor representations \(\mathbf{10}_a,\ \mathbf{10^*}_a,\ \mathbf{8}_a\) combine with the space-spin symmetric partial waves \({}^3S_1,\ {}^1P_1,\ {}^3D_1\leftrightarrow{}^3S_1\) with the 12 constants \(\tilde c_{{}^3S_1}^{8a,10,10^*}\), \(c_{{}^3S_1}^{8a,10,10^*}\), \(c_{{}^1P_1}^{8a,10,10^*}\) and \(c_{{}^3D_1\text-{}^3S_1}^{8a,10,10^*}\).
Additionally, as stated before, the Lagrangian \#28 in Table~\ref{tab:BBLagrangians} leads to the spin singlet-triplet transitions \({}^1P_1\leftrightarrow{}^3P_1\).
The corresponding low-energy constant for this transition between the \(\mathbf 8_a\) and \(\mathbf 8_s\) representations is \(c^{8as}\).
The constants \(\tilde c_{{}^1S_0}^{27,8s,1}\) and \(\tilde c_{{}^3S_1}^{8a,10,10^*}\) fulfill the same SU(3) relations as the constants \(c_{{}^1S_0}^{27,8s,1}\) and \(c_{{}^3S_1}^{8a,10,10^*}\) in Table~\ref{tab:PWDBB}.
The SU(3) breaking linear in the quark masses appears only in the S-waves \({}^1S_0,\ {}^3S_1\), and is proportional \(m_K^2-m_\pi^2\). The corresponding 12 constants are \(c_\chi^{1,\dots,12}\).
The SU(3) symmetric relations can also be derived by group theoretical
considerations. For the SU(3) breaking part this is not feasible and the
contributions have to be derived from the Lagrangian.

In order to obtain the complete partial wave projected potentials, the entries in Table~\ref{tab:PWDBB} have to be multiplied with additional factors.
The leading order constants \(\tilde c^{\,i}_j\) do not need to be multiplied with a factor. For the next-to-leading order constants (without tilde and without \(\chi\)) the partial waves \({}^1S_0,\ {}^3S_1\) have to be multiplied with a factor  \((p^2_i+p_f^2)\).
The contribution to the partial waves \({}^1S_0,\ {}^3S_1\) from constants \(c^j_\chi\) has to be multiplied with \((m_K^2-m_\pi^2)\).
The partial waves \({}^3P_0,\ {}^3P_1,\ {}^3P_2,\ {}^1P_1,\ {}^1P_1\leftrightarrow{}^3P_1\) get multiplied with \(p_ip_f\).
The entries for \({}^3S_1\rightarrow{}^3D_1\) and \({}^3D_1\rightarrow{}^3S_1\) have to be multiplied with \(p_i^2\) and \(p_f^2\), respectively.
One obtains for example for the \(NN\) interaction in the \({}^1S_0\) partial wave:
\begin{equation}
 \langle NN, {}^1S_0|\hat V|NN, {}^1S_0\rangle = \tilde c^{27}_{{}^1S_0} + c^{27}_{{}^1S_0} (p^2_i+p_f^2) + \frac12 c^1_\chi(m_K^2-m_\pi^2)  \,,
\end{equation}
or for the \(\Xi N\rightarrow\Sigma\Sigma\) interaction with total isospin \(I=0\) in the \({}^1P_1\rightarrow{}^3P_1\) partial wave:
\begin{equation}
 \langle \Sigma\Sigma, {}^3P_1|\hat V|\Xi N, {}^1P_1\rangle = 2\sqrt3 c^{8as} p_ip_f \,.
\end{equation} 
When restricting to the \(NN\) channel we recover the well-known two leading and seven next-to-leading order low-energy constants of Ref.~\cite{Epelbaum2004} in the different partial wave channels (\({}^1S_0\), \({}^3S_1\), \({}^1P_1\), \({}^3P_{0}\), \({}^3P_{1}\), \({}^3P_{2}\), \({}^3S_1\leftrightarrow {}^3D_1 \)).

\section{Summary and conclusions} \label{sec:summary}

In this work we have constructed the relativistically invariant chiral baryon-baryon contact Lagrangian in flavor SU(3) up to order \(\mathcal{O}(q^2)\).
It provides the contact terms for baryon-baryon scattering consistent with SU(3) flavor symmetry (and the discrete and Lorentz symmetries of the strong interactions).
By employing the external field method a full set of baryon-baryon contact terms involving pseudoscalar mesons and/or external electroweak fields has been constructed.

For pure baryon-baryon contact terms we have presented a minimal set of 40 chiral contact Lagrangians up to \(\mathcal O (q^2)\).
These contact potentials are important for the description of hyperon-nucleon and hyperon-hyperon scattering processes.
When decomposed into partial waves, 28 of these contact terms lead to SU(3) symmetric contributions to the potentials for the channels \({}^1S_0\), \({}^3S_1\), \({}^1P_1\), \({}^3P_0\), \({}^3P_1\), \({}^3P_2\), \({}^3D_1\leftrightarrow{}^3S_1\) and \({}^1P_1\leftrightarrow{}^3P_1\).
In particular we have identified the specific contact term  of the chiral Lagrangian which provides the antisymmetric spin-orbit interaction.
The remaining 12 low-energy constants contribute to the \({}^1S_0\) and \({}^3S_1\) partial wave and lead to SU(3) symmetry breaking contributions linear in the quark masses.

The contact terms involving pseudoscalar Goldstone-boson fields and/or electroweak fields come into play in the description of chiral many-body forces and exchange-currents relevant for few-baryon systems.

\section{Acknowledgement}

We thank J.~Haidenbauer and W.~Weise for useful discussions.
This work is supported in part by the DFG and the NSFC through funds provided to the Sino-German CRC 110 ``Symmetries and the Emergence of Structure in QCD''.
S.~Petschauer thanks the ``TUM Graduate School''.

\begin{table}[htb]
\centering
\begin{tabular}{>{\(}c<{\)}>{\(}c<{\)}>{\(}c<{\)}}
\toprule
 \# & \mathscr L & \text{contributes to} \\
\midrule
 1 & \trace{\bar B_1 B_1 \bar B_2 B_2} & P_1,P_4 \\
 2 & \trace{\bar B_1(\partial^\mu B)_1 \bar B_2(\partial_\mu B)_2} + \trace{ (\partial^\mu \bar B)_1 B_1 (\partial_\mu \bar B)_2 B_2} & P_1,P_4 \\
 3 & \trace{\bar B_1(\gamma_5\gamma^\mu B)_1 \bar B_2(\gamma_5\gamma_\mu B)_2} & P_2,P_3,P_4,P_9 \\
 4 & \trace{\bar B_1(\gamma_5\gamma^\mu\partial^\nu B)_1 \bar B_2(\gamma_5\gamma_\mu \partial_\nu B)_2} + \trace{ (\partial^\nu \bar B)_1 (\gamma_5\gamma^\mu B)_1 (\partial_\nu \bar B)_2 (\gamma_5\gamma_\mu B)_2} & P_2,P_3,P_4,P_9 \\
 5 & \trace{\bar B_1(\gamma_5\gamma^\mu\partial^\nu B)_1 \bar B_2(\gamma_5\gamma_\nu \partial_\mu B)_2} + \trace{ (\partial^\nu \bar B)_1 (\gamma_5\gamma^\mu B)_1 (\partial_\mu \bar B)_2 (\gamma_5\gamma_\nu B)_2} & P_2,P_3,P_9 \\
 6 & \trace{\bar B_1(\sigma^{\mu\nu} B)_1 \bar B_2(\sigma_{\mu\nu} B)_2} & P_1,P_2,P_3,P_4,P_9 \\
 7 & \hat\partial^2_2 \trace{\bar B_1 B_1 \bar B_2 B_2} & P_1 \\
 8 & \hat\partial^2_2 \trace{\bar B_1(\gamma_5\gamma^\mu B)_1 \bar  B_2(\gamma_5\gamma_\mu B)_2} & P_2 \\
 9 & \hat\partial_2^\alpha\hat\partial_2^\beta\trace{\bar B_1(\gamma_5\gamma_\alpha B)_1 \bar B_2(\gamma_5\gamma_\beta B)_2} & P_2,P_3 \\[1.5ex]
 10 & \trace{\bar B_1 \bar B_2 B_1 B_2} & P_1,P_4 \\
 11 & \trace{\bar B_1\bar B_2(\partial^\mu B)_1 (\partial_\mu B)_2} + \trace{ (\partial^\mu \bar B)_1 (\partial_\mu \bar B)_2 B_1 B_2} & P_1,P_4 \\
 12 & \trace{\bar B_1\bar B_2(\gamma_5\gamma^\mu B)_1 (\gamma_5\gamma_\mu B)_2} & P_2,P_3,P_4,P_9 \\
 13 & \trace{\bar B_1\bar B_2(\gamma_5\gamma^\mu\partial^\nu B)_1 (\gamma_5\gamma_\mu \partial_\nu B)_2} + \trace{ (\partial^\nu \bar B)_1 (\partial_\nu \bar B)_2 (\gamma_5\gamma^\mu B)_1 (\gamma_5\gamma_\mu B)_2} & P_2,P_3,P_4,P_9 \\
 14 & \trace{\bar B_1\bar B_2(\gamma_5\gamma^\mu\partial^\nu B)_1 (\gamma_5\gamma_\nu \partial_\mu B)_2} + \trace{ (\partial^\nu \bar B)_1 (\partial_\mu \bar B)_2 (\gamma_5\gamma^\mu B)_1 (\gamma_5\gamma_\nu B)_2} & P_2,P_3,P_9 \\
 15 & \trace{\bar B_1\bar B_2(\sigma^{\mu\nu} B)_1 (\sigma_{\mu\nu} B)_2} & P_1,P_2,P_3,P_4,P_9 \\
 16 & (\hat\partial^2_2+\hat\partial^2_1) \trace{\bar B_1 \bar B_2B_1  B_2} & P_1 \\
 17 & (\hat\partial^2_2+\hat\partial^2_1) \trace{\bar B_1\bar  B_2(\gamma_5\gamma^\mu B)_1 (\gamma_5\gamma_\mu B)_2} & P_2 \\
 18 & (\hat\partial_2^\alpha\hat\partial_2^\beta+\hat\partial_1^\alpha\hat\partial_1^\beta) \trace{\bar B_1\bar B_2(\gamma_5\gamma_\alpha B)_1 (\gamma_5\gamma_\beta B)_2} & P_2,P_3 \\[1.5ex]
 19 & \trace{\bar B_1 B_1 }\trace{ \bar B_2 B_2} & P_1,P_4 \\
 20 & \trace{\bar B_1(\partial^\mu B)_1 }\trace{ \bar B_2(\partial_\mu B)_2} + \trace{ (\partial^\mu \bar B)_1 B_1 }\trace{ (\partial_\mu \bar B)_2 B_2} & P_1,P_4 \\
 21 & \trace{\bar B_1(\gamma_5\gamma^\mu B)_1 }\trace{ \bar B_2(\gamma_5\gamma_\mu B)_2} & P_2,P_3,P_4,P_9 \\
 22 & \trace{\bar B_1(\gamma_5\gamma^\mu\partial^\nu B)_1 }\trace{ \bar B_2(\gamma_5\gamma_\mu \partial_\nu B)_2} + \trace{ (\partial^\nu \bar B)_1 (\gamma_5\gamma^\mu B)_1 }\trace{ (\partial_\nu \bar B)_2 (\gamma_5\gamma_\mu B)_2} & P_2,P_3,P_4,P_9 \\
 23 & \trace{\bar B_1(\gamma_5\gamma^\mu\partial^\nu B)_1 }\trace{ \bar B_2(\gamma_5\gamma_\nu \partial_\mu B)_2} + \trace{ (\partial^\nu \bar B)_1 (\gamma_5\gamma^\mu B)_1 }\trace{ (\partial_\mu \bar B)_2 (\gamma_5\gamma_\nu B)_2} & P_2,P_3,P_9 \\
 24 & \trace{\bar B_1(\sigma^{\mu\nu} B)_1 }\trace{ \bar B_2(\sigma_{\mu\nu} B)_2} & P_1,P_2,P_3,P_4,P_9 \\
 25 & \hat\partial^2_2 \trace{\bar B_1 B_1 } \trace{ \bar B_2 B_2} & P_1 \\
 26 & \hat\partial^2_2 \trace{\bar B_1(\gamma_5\gamma^\mu B)_1 } \trace{ \bar  B_2(\gamma_5\gamma_\mu B)_2} & P_2 \\
 27 & \hat\partial_2^\alpha\hat\partial_2^\beta \trace{\bar B_1(\gamma_5\gamma_\alpha B)_1 } \trace{ \bar B_2(\gamma_5\gamma_\beta B)_2} & P_2,P_3 \\[1.5ex]
 28 & \hat \partial_2^\alpha\trace{\bar B_1\bar B_2(\gamma_5\gamma_\alpha \partial_\mu B)_1(\gamma_5\gamma^\mu B)_2} + \hat \partial_1^\alpha\trace{\bar B_1 (\partial_\mu\bar B)_2 (\gamma_5\gamma^\mu B)_1(\gamma_5\gamma_\alpha B)_2} & P_2,P_3,P_8\\[1.5ex]
 29 & \trace{\bar B_1 \chi B_1\bar B_2 B_2} & P_1 \\
 30 & \trace{\bar B_1 \chi(\gamma_5\gamma^\mu B)_1\bar B_2(\gamma_5\gamma_\mu B)_2} & P_2 \\
 31 & \trace{\bar B_1 B_1 \chi\bar B_2 B_2} & P_1 \\
 32 & \trace{\bar B_1 (\gamma_5\gamma^\mu B)_1 \chi\bar B_2 (\gamma_5\gamma_\mu B)_2} & P_2 \\
 33 & \trace{\bar B_1\chi\bar B_2B_1B_2} + \trace{\bar B_1\bar B_2B_1\chi B_2} & P_1 \\
 34 & \trace{\bar B_1\chi\bar B_2(\gamma_5\gamma^\mu B)_1(\gamma_5\gamma_\mu B)_2} + \trace{\bar B_1\bar B_2(\gamma_5\gamma^\mu B)_1\chi (\gamma_5\gamma_\mu B)_2} & P_2 \\
 35 & \trace{\bar B_1\bar B_2\chi B_1B_2} & P_1 \\
 36 & \trace{\bar B_1\bar B_2\chi (\gamma_5\gamma^\mu B)_1(\gamma_5\gamma_\mu B)_2} & P_2 \\
 37 & \trace{\bar B_1\bar B_2B_1B_2\chi} & P_1 \\
 38 & \trace{\bar B_1\bar B_2(\gamma_5\gamma^\mu B)_1(\gamma_5\gamma_\mu B)_2\chi} & P_2 \\
 39 & \trace{\bar B_1\chi B_1}\trace{\bar B_2B_2} & P_1 \\
 40 & \trace{\bar B_1\chi (\gamma_5\gamma^\mu B)_1}\trace{\bar B_2(\gamma_5\gamma_\mu B)_2} & P_2 \\
\bottomrule
\end{tabular}
\caption{Linearly independent Lagrangians up to \(\mathcal O(q^2)\) for pure baryon-baryon interaction in non-relativistic power counting and their contribution in spin-space.} \label{tab:BBLagrangians}
\end{table}

\newgeometry{left=.7cm,right=.7cm, top=.7cm, bottom=.7cm}
\begin{sidewaystable}
\centering
\resizebox*{.99\textheight}{!}{
\begin{tabular}{>{\(}c<{\)}>{\(}c<{\)}>{\(}c<{\)}>{\(}c<{\)}>{\(}c<{\)}>{\(}c<{\)}>{\(}c<{\)}>{\(}c<{\)}>{\(}c<{\)}}
\toprule
 S & I & \text{transition} & j\in\{{}^1S_0, {}^3P_0, {}^3P_1, {}^3P_2\} & j\in\{{}^3S_1, {}^1P_1, {}^3S_1\leftrightarrow{}^3D_1\} & {}^1P_1\rightarrow {}^3P_1 & {}^3P_1\rightarrow {}^1P_1 & {}^1S_0 \ \chi & {}^3S_1 \ \chi \\
\cmidrule(lr){1-3}\cmidrule(lr){4-9}
  0 & 0 & NN\rightarrow NN & 0 & c^{10^*}_{j} & 0 & 0 & 0 & \frac{c_\chi^7}{2} \\
   & 1 & NN\rightarrow NN & c^{27}_{j} & 0 & 0 & 0 & \frac{c_\chi^1}{2} & 0 \\
\cmidrule(lr){1-3}\cmidrule(lr){4-9}
 -1  & \frac12 & \Lambda N\rightarrow \Lambda N & \frac{1}{10} (9 c^{27}_{j}+c^{8s}_{j}) & \frac{1}{2}(c^{10^*}_{j}+c^{8a}_{j}) & -c^{8as} & -c^{8as} & c_\chi^2 & c_\chi^8 \\
  & \frac12 & \Lambda N\rightarrow \Sigma N & -\frac{3}{10} (c^{27}_{j}-c^{8s}_{j}) & \frac{1}{2}(c^{10^*}_{j}-c^{8a}_{j}) & -3 c^{8as} & c^{8as} & -c_\chi^3 & -c_\chi^9 \\
  & \frac12 & \Sigma N\rightarrow \Sigma N & \frac{1}{10} (c^{27}_{j}+9 c^{8s}_{j}) & \frac{1}{2}(c^{10^*}_{j}+c^{8a}_{j}) & 3 c^{8as} & 3 c^{8as} & c_\chi^4 & c_\chi^{10} \\
  & \frac32 & \Sigma N\rightarrow \Sigma N & c^{27}_{j} & c^{10}_{j} & 0 & 0 & \frac{c_\chi^1}{4} & -\frac{c_\chi^7}{4} \\
\cmidrule(lr){1-3}\cmidrule(lr){4-9}
  -2 & 0 & \Lambda \Lambda \rightarrow \Lambda \Lambda & \frac{1}{40} (5 c^{1}_{j}+27 c^{27}_{j}+8 c^{8s}_{j}) & 0 & 0 & 0 & \frac{c_\chi^5}{2} & 0 \\
  & 0 & \Lambda \Lambda \rightarrow \Xi N & \frac{1}{20} (5 c^{1}_{j}-9 c^{27}_{j}+4 c^{8s}_{j}) & 0 & 0 & 2 c^{8as} & \frac{3 c_\chi^1}{4}-3 c_\chi^2-c_\chi^3+\frac{3 c_\chi^5}{4} & 0 \\
  & 0 & \Lambda \Lambda \rightarrow \Sigma \Sigma & -\frac{\sqrt{3}}{40}  (5 c^{1}_{j}+3 c^{27}_{j}-8 c^{8s}_{j}) & 0 & 0 & 0 & 0 & 0 \\
  & 0 & \Xi N \rightarrow \Xi N & \frac{1}{10} (5 c^{1}_{j}+3 c^{27}_{j}+2 c^{8s}_{j}) & c^{8a}_{j} & 2 c^{8as} & 2 c^{8as} & \frac{2 c_\chi^1}{3}-3 c_\chi^2+\frac{c_\chi^4}{3}+\frac{9 c_\chi^5}{8} & c_\chi^{11} \\
  & 0 & \Xi N \rightarrow \Sigma \Sigma & \frac{\sqrt{3}}{20}  (-5 c^{1}_{j}+c^{27}_{j}+4 c^{8s}_{j}) & 0 & 2 \sqrt{3} c^{8as} & 0 & -\frac{c_\chi^1}{4 \sqrt{3}}+\sqrt{3} c_\chi^3+\frac{c_\chi^4}{\sqrt{3}} & 0 \\
  & 0 & \Sigma \Sigma \rightarrow \Sigma \Sigma & \frac{1}{40} (15 c^{1}_{j}+c^{27}_{j}+24 c^{8s}_{j}) & 0 & 0 & 0 & 0 & 0 \\
  & 1 & \Xi N \rightarrow \Xi N & \frac{1}{5} (2 c^{27}_{j}+3 c^{8s}_{j}) & \frac{1}{3} (c^{10}_{j}+c^{10^*}_{j}+c^{8a}_{j}) & -2 c^{8as} & -2 c^{8as} & c_\chi^6 & c_\chi^{12} \\
  & 1 & \Xi N \rightarrow \Sigma \Sigma  & 0 & \frac{1}{3 \sqrt{2}}(c^{10}_{j}+c^{10^*}_{j}-2 c^{8a}_{j}) & 0 & 2 \sqrt{2} c^{8as} & 0 & \sqrt{2} c_\chi^{10}-\frac{c_\chi^7}{2 \sqrt{2}}-\sqrt{2} c_\chi^9 \\
  & 1 & \Xi N \rightarrow \Sigma \Lambda  & \frac{\sqrt{6}}{5}  (c^{27}_{j}-c^{8s}_{j}) & \frac{1}{\sqrt{6}}(c^{10}_{j}-c^{10^*}_{j}) & 2 \sqrt{\frac{2}{3}} c^{8as} & 0 & -\frac{1}{3} \sqrt{\frac{2}{3}} c_\chi^1+\sqrt{\frac{3}{2}} c_\chi^2-\frac{c_\chi^4}{3 \sqrt{6}}-\sqrt{\frac{2}{3}} c_\chi^6 & \frac{c_\chi^{10}}{\sqrt{6}}+\sqrt{\frac{2}{3}} c_\chi^{12}+\frac{c_\chi^7}{2 \sqrt{6}}-\sqrt{\frac{3}{2}} c_\chi^8+\sqrt{\frac{2}{3}} c_\chi^9 \\
  & 1 & \Sigma \Lambda \rightarrow \Sigma \Lambda  & \frac{1}{5} (3 c^{27}_{j}+2 c^{8s}_{j}) & \frac{1}{2}(c^{10}_{j}+c^{10^*}_{j}) & 0 & 0 & -\frac{c_\chi^1}{9}+\frac{4 c_\chi^3}{3}+\frac{4 c_\chi^4}{9}+\frac{2 c_\chi^6}{3} & \frac{4 c_\chi^{10}}{3}+\frac{2 c_\chi^{12}}{3}-\frac{c_\chi^7}{3}-\frac{4 c_\chi^9}{3} \\
  & 1 & \Sigma \Lambda \rightarrow \Sigma \Sigma  & 0 & \frac{1}{2 \sqrt{3}}(c^{10}_{j}-c^{10^*}_{j}) & 0 & \frac{4}{\sqrt{3}} c^{8as} & 0 & 0 \\
  & 1 & \Sigma \Sigma \rightarrow \Sigma \Sigma  & 0 & \frac{1}{6} (c^{10}_{j}+c^{10^*}_{j}+4 c^{8a}_{j}) & 0 & 0 & 0 & 0 \\
  & 2 & \Sigma \Sigma \rightarrow \Sigma \Sigma  & c^{27}_{j} & 0 & 0 & 0 & 0 & 0 \\
\cmidrule(lr){1-3}\cmidrule(lr){4-9}
  -3 & \frac12 & \Xi \Lambda \rightarrow \Xi \Lambda & \frac{1}{10} (9 c^{27}_{j}+c^{8s}_{j}) & \frac{1}{2}(c^{10}_{j}+c^{8a}_{j}) & -c^{8as} & -c^{8as} & -\frac{55 c_\chi^1}{72}+2 c_\chi^2+\frac{7 c_\chi^3}{6}+\frac{c_\chi^4}{18}+\frac{3 c_\chi^5}{32}+\frac{c_\chi^6}{12} & \frac{11 c_\chi^{10}}{12}+\frac{3 c_\chi^{11}}{4}+\frac{25 c_\chi^{12}}{12}+\frac{5 c_\chi^7}{24}-\frac{7 c_\chi^8}{4}-\frac{c_\chi^9}{6} \\
  & \frac12 & \Xi \Lambda \rightarrow \Xi \Sigma & -\frac{3}{10} (c^{27}_{j}-c^{8s}_{j}) & \frac{1}{2}(c^{10}_{j}-c^{8a}_{j}) & -3 c^{8as} & c^{8as} & \frac{11 c_\chi^1}{24}-\frac{3 c_\chi^2}{2}-\frac{c_\chi^3}{2}-\frac{c_\chi^4}{3}+\frac{9 c_\chi^5}{32}+\frac{c_\chi^6}{4} & \frac{9 c_\chi^{10}}{4}-\frac{3 c_\chi^{11}}{4}+\frac{5 c_\chi^{12}}{4}-\frac{c_\chi^7}{8}-\frac{3 c_\chi^8}{4}-\frac{c_\chi^9}{2} \\
  & \frac12 & \Xi \Sigma \rightarrow \Xi \Sigma & \frac{1}{10} (c^{27}_{j}+9 c^{8s}_{j}) & \frac{1}{2}(c^{10}_{j}+c^{8a}_{j}) & 3 c^{8as} & 3 c^{8as} & \frac{11 c_\chi^1}{24}-3 c_\chi^2+\frac{5 c_\chi^3}{2}+\frac{c_\chi^4}{6}+\frac{27 c_\chi^5}{32}+\frac{3 c_\chi^6}{4} & \frac{5 c_\chi^{10}}{4}+\frac{3 c_\chi^{11}}{4}+\frac{3 c_\chi^{12}}{4}-\frac{c_\chi^7}{8}-\frac{3 c_\chi^8}{4}-\frac{3 c_\chi^9}{2} \\
  & \frac32 & \Xi \Sigma \rightarrow \Xi \Sigma & c^{27}_{j} & c^{10^*}_{j} & 0 & 0 & -\frac{2 c_\chi^1}{3}+\frac{3 c_\chi^2}{2}+c_\chi^3+\frac{c_\chi^4}{6} & \frac{3 c_\chi^{10}}{2}-c_\chi^7+\frac{3 c_\chi^8}{2}-3 c_\chi^9 \\
\cmidrule(lr){1-3}\cmidrule(lr){4-9}
  -4 & 0 & \Xi \Xi \rightarrow \Xi \Xi & 0 & c^{10}_{j} & 0 & 0 & 0 & 5 c_\chi^{10}+4 c_\chi^{12}-3 c_\chi^8-2 c_\chi^9 \\
  & 1 & \Xi \Xi \rightarrow \Xi \Xi & c^{27}_{j} & 0 & 0 & 0 & -\frac{4 c_\chi^1}{3}+3 c_\chi^2+2 c_\chi^3+\frac{c_\chi^4}{3} & 0 \\
\bottomrule
\end{tabular}
}
\caption{
SU(3) relations of pure baryon-baryon contact terms in non-vanishing partial waves up to \(\mathcal O(q^2)\) with non-relativistic power counting.} \label{tab:PWDBB}
\end{sidewaystable}
\restoregeometry

\clearpage

\appendix

\section{Reduction by using the baryon equation of motion} \label{sec:EOM}

In order to avoid the construction of redundant terms, one uses the equation of motion fulfilled by the baryon field
\begin{equation}
 \slashed{D}B = \gamma^\mu D_\mu B = -\mathrm i M_0 B + \mathcal{O}(q) \,,
\end{equation}
and its Dirac conjugated. Up to higher order corrections one can replace \(\slashed{D}B\) by \(-\mathrm i M_0 B\) and \(\bar B \overleftarrow{\slashed{D}}\) by \(\mathrm i M_0 \bar B\).
Beyond the obvious replacements one can bring terms not containing \(\slashed{D}B\) into a form where they do.
We follow closely Ref.~\cite{Fettes2000}, where this method has been applied to the \(\pi N\) Lagrangian.
The first step in the reduction is to rewrite a product of a Dirac matrix with \(\gamma_\lambda\) as a sum of other Dirac matrices multiplied with factors of \(g_{\mu\nu}\) and \(\epsilon_{\mu\nu\rho\tau}\)-tensors, sorted by their behavior under charge conjugation:
\begin{equation} \label{eq:gamma-rel}
 \Gamma\gamma_\lambda = \Gamma^\prime_{\lambda} + \Gamma^{\prime\prime}_{\lambda}\,,\quad c_{\Gamma^{\prime\prime}}=c_{\Gamma^\prime}+1=c_\Gamma\,.
\end{equation}
Table~\ref{tab:gammadec} gives the result of these gamma-matrix decompositions for the basis elements \(\mathbbm 1\), \(\gamma_\mu\), \(\gamma_5\), \(\gamma_5\gamma_\mu\), \(\sigma_{\mu\nu}\) and their contraction with the \(\epsilon\)-tensor.

\begin{table}[htb!]
\centering
\begin{tabular}{>{\(}c<{\)}>{\(}c<{\)}>{\(}c<{\)}>{\(}c<{\)}>{\(}l<{\)}}
 \toprule
 \Gamma & \Gamma^\prime_{\lambda} & \Gamma^{\prime\prime}_{\lambda} \\
 \midrule
 \mathbbm{1} & \gamma_\lambda & 0 \\
 \gamma_\mu & g_{\mu\lambda} \mathbbm{1} & -\mathrm i \sigma_{\mu\lambda} \\
 \gamma_5 & 0 & \gamma_5\gamma_\lambda \\
 \gamma_5\gamma_\mu & \frac12 \epsilon_{\mu\lambda\rho\tau}\sigma^{\rho\tau} & g_{\mu\lambda} \gamma_5 \\
 \sigma_{\mu\nu} & \epsilon_{\mu\nu\lambda\tau}\gamma_5\gamma^\tau & -\mathrm i (g_{\mu\lambda}\gamma_\nu-g_{\nu\lambda}\gamma_\mu) \\
 \midrule
 \epsilon_{\mu\nu\rho\tau}\gamma^\tau & \epsilon_{\mu\nu\rho\lambda} \mathbbm{1} & g_{\mu\lambda}\gamma_5\sigma_{\nu\rho} + g_{\rho\lambda}\gamma_5\sigma_{\mu\nu} + g_{\nu\lambda}\gamma_5\sigma_{\rho\mu} \\
 \epsilon_{\mu\nu\rho\tau}\gamma_5\gamma^\tau & g_{\mu\lambda}\sigma_{\nu\rho} + g_{\rho\lambda}\sigma_{\mu\nu} + g_{\nu\lambda}\sigma_{\rho\mu} & \epsilon_{\mu\nu\rho\lambda}\gamma_5 \\
 \epsilon_{\mu\nu\rho\alpha}{\sigma^{\alpha}}_\tau & \gamma_5\gamma_\rho(g_{\lambda\nu}{g_{\mu\tau}} - g_{\lambda\mu}{g_{\nu\tau}}) & \mathrm i {g_{\lambda\tau}}\epsilon_{\mu\nu\rho\alpha}\gamma^\alpha - \mathrm i \epsilon_{\mu\nu\rho\lambda}\gamma_\tau\\
 & + \gamma_5\gamma_\nu(g_{\lambda\mu}{g_{\rho\tau}} - g_{\lambda\rho}{g_{\mu\tau}}) & \\
 & + \gamma_5\gamma_\mu(g_{\lambda\rho}{g_{\nu\tau}} - g_{\lambda\nu}{g_{\rho\tau}}) & \\
 \frac{\mathrm i}{2} \epsilon_{\mu\nu\rho\tau}\sigma^{\rho\tau} = \gamma_5\sigma_{\mu\nu} & \frac{1}{\mathrm i}(g_{\mu\lambda}\gamma_5\gamma_\nu-g_{\nu\lambda}\gamma_5\gamma_\mu) & \epsilon_{\mu\nu\lambda\rho}\gamma^\rho \\
 \bottomrule
\end{tabular}
\caption{Decomposition of the Dirac matrix product \(\Gamma\gamma_\lambda\) into charge conjugation even and charge conjugation odd parts.} \label{tab:gammadec}
\end{table}
Now one plugs these matrices into a monomial with \(D^\lambda\), and obtains a relation where \(\slashed{D}B\) can be replaced by \(-\mathrm i M_0 B\).
In order to exploit the charge conjugation behavior we define for a monomial \(X\) the combination \(Y\) as
\begin{equation} \label{eq:defy}
 Y \defeq X \pm (-1)^{c+n}X^C\,,
\end{equation}
where \(c\) is the charge conjugation exponent of the monomial \(X\) and \(n=n_1+n_2\) is the number of covariant derivatives occurring in the two Dirac operators of the monomial \(X\).
The combination \(Y\) transforms under charge conjugation as
\begin{equation}
 Y^C = \pm (-1)^{c+n} Y \,,
\end{equation}
where the plus or minus sign can be chosen in Eq.~\eqref{eq:defy} such that the term \(Y\) becomes charge conjugation invariant.
Notice that a modification of the Dirac operators \(\Theta^i\) in the monomial \(X\) leads to the same modification of the Dirac operators \(\Theta^i\) in the term \((-1)^{c+n}X^C\).
If there occurs a sign change in the charge-conjugate of the modified term \(X^C\), it gets compensated by the factor \((-1)^{c+n}\).

Now one can insert the Dirac matrix decomposition of Eq.~\eqref{eq:gamma-rel} into the combination \(Y\) and obtains, after applying the equation of motion,
\begin{equation} \label{eq:proof1}
 -\mathrm iM_0 \underbrace{Y(\Theta^i = \Gamma D^{n_i-1})}_{c\,,\ n-1}
 \apr Y(\Theta^i = \Gamma\gamma^\lambda D^{n_i}_{\lambda}) \\
 = \underbrace{Y(\Theta^i = {\Gamma^\prime}^\lambda D^{n_i}_{\lambda})}_{c+1\,,\ n}
 + \underbrace{Y(\Theta^i = {\Gamma^{\prime\prime}}^\lambda D^{n_i}_{\lambda})}_{c\,,\ n} \,,
\end{equation}
where we have indicated the charge conjugation exponent \(c\) and the number of covariant derivatives \(n=n_1+n_2\) under each term.
The symbol \(\apr\) means that both sides are equal up to higher order terms.
Taking the charge conjugate of both sides of Eq.~\eqref{eq:proof1} leads to
\begin{equation}
 \label{eq:proof2}
 -\mathrm iM_0  (-1)^{c+n-1}Y(\Theta^i = \Gamma D^{n_i-1}) \\
 \apr (-1)^{c+n+1} Y(\Theta^i = {\Gamma^\prime}^\lambda D^{n_i}_{\lambda})
 + (-1)^{c+n} Y(\theta_i = {\Gamma^{\prime\prime}}^\lambda D^{n_i}_{\lambda})\,.
\end{equation}
Solving the system of two linear equations \eqref{eq:proof1} and \eqref{eq:proof2}, one gets two relations for charge conjugation invariant terms:
\begin{equation} \begin{aligned} \label{eq:eomrel}
 Y(\Theta^i = {\Gamma^\prime}^\lambda D^{n_i}_{\lambda}) &\apr -\mathrm i M_0 Y(\Theta^i = {\Gamma} D^{n_i-1})\,, \\
 Y(\Theta^i = {\Gamma^{\prime\prime}}^\lambda D^{n_i}_{\lambda}) &\apr 0\,.
\end{aligned} \end{equation}

Using this result, together with the decomposition of the Dirac matrix products in Table~\ref{tab:gammadec} and the relation \(Y(\Theta^i = \dots D_\mu D^\mu) \apr - M_0^2 Y(\Theta^i = \dots)\), gives rise to various equations, which can be summarized by the following restrictions for remaining independent terms \cite{Fettes2000}:
\begin{itemize}
 \item \(\Gamma\) of \(\Theta^i\) is a matrix from the set \(\{\mathbbm{1},\gamma_5\gamma_\mu,\sigma_{\mu\nu}\}\) with possible additional \(g_{\mu\nu}\) and \(\epsilon_{\mu\nu\rho\tau}\) factors,
 \item Lorentz indices within \(\Theta^i\) must not be contracted, with the exception of one index of an \(\epsilon_{\mu\nu\rho\tau}\)-tensor, which has then to be contracted with \(D^n_{\mu\dots}\,\).
\end{itemize}

\bigskip
The equations of motion restrict also the Lorentz contractions concerning the covariant derivative \(\hat D_2^\mu\) of a baryon bilinear.
One obtains for a charge conjugation invariant monomial \(Y\), using Eq.~\eqref{eq:eomrel} and repeatedly the equations of motion, the following relations
\begin{equation} \begin{aligned} \label{eq:dmucontr}
 &\hat D_2^\mu Y(\Theta^2 = \mathbbm{1} D^n_{\mu\dots}) \apr 0 \,,\\
 &\hat D_2^\mu Y(\Theta^2 = \gamma_5 D^n_{\mu\dots}) \apr 0 \,,\\
 &\hat D_2^\mu Y(\Theta^2 = \gamma_\mu D^n) \apr 0 \,,\\
 &\hat D_2^\mu Y(\Theta^2 = \gamma_\nu D^n_{\mu\dots}) \apr 0 \,,\\
 &\hat D_2^\mu Y(\Theta^2 = \gamma_5\gamma_\mu D^n) \apr -2\mathrm i M_0 Y(\theta_2 = \gamma_5 D^n)\,,\\
 &\hat D_2^\mu Y(\Theta^2 = \gamma_5\gamma_\nu D^n_{\mu\dots}) \apr 0 \,,\\
 &\hat D_2^\mu Y(\Theta^2 = {\sigma_\mu}^{\nu} D^n) \apr -\mathrm i \hat D_2^\nu Y(\theta_2 =  D^n)\,,\\
 &\hat D_2^\mu Y(\Theta^2 = \sigma_{\nu\rho} D^n_{\mu\dots}) \apr 0\,.
\end{aligned} \end{equation}
For contractions with \(\Theta^1\) the same rules hold, but with an additional minus sign, since covariant derivatives of baryon bilinears \(\hat D_2^\mu\) can be replaced by \(-\hat D_1^\mu\), up to terms of higher order or already included, see Eq.~\eqref{eq:totalder}.
The set of relations in \eqref{eq:dmucontr} leads to the following additional rule for independent terms
\begin{itemize}
 \item \(\hat D_2^\mu\) must not be contracted with any \(D^n_{\mu\dots}\) or with \(\sigma_{\mu\nu}\).
\end{itemize}

\section{Non-relativistic expansion} \label{sec:HBexp}

Via the Feynman rules the baryon-baryon contact terms get translated into matrix elements of the form \((\bar u_3\Gamma_1 u_1)(\bar u_4 \Gamma_2 u_2)\) with additional momentum factors from derivatives.
A derivative acting on a baryon field, \(D_\mu B\), is counted of order \(\mathcal O (q^0)\) for \(\mu=0\) and of order \(\mathcal O (q^1)\) for \(\mu=1,2,3\).
The free Dirac spinors have the well-known form
\begin{equation}
 \bar u_i = \sqrt{\frac{E_i+M_0}{2M_0}}\left( \mathbbm1 \ ,\ -\frac{\vec\sigma\cdot\vec p^{\,\prime}}{E_i+M_0} \right)\,,\quad
 u_j = \sqrt{\frac{E_j+M_0}{2M_0}}\begin{pmatrix} \mathbbm1 \\ \frac{\vec\sigma\cdot\vec p}{E_j+M_0} \end{pmatrix} \,,
\end{equation}
with
\begin{equation}
 E_i = \sqrt{M_0^2 + \vec p^{\,\prime2}}\,,\quad
 E_j = \sqrt{M_0^2 + \vec p^{\,2}}\,,
\end{equation}
where \(M_0\) denotes a common baryon mass in the chiral limit.
If one expands the baryon bilinears in the inverse large baryon mass, one obtains the following expressions up to order \(\mathcal O (q^2)\):
\begin{equation} \begin{aligned}
 \bar u_i \mathbbm1 u_j &\approx 1\ +\ \frac{p^2+p^{\prime\,2}}{8M_0^2} - \frac{\vec \sigma\cdot\vec p^{\,\prime} \ \vec \sigma\cdot\vec p}{4M_0^2} \,,\\
 \bar u_i \gamma^0 u_j &\approx 1 \ +\ \frac{p^2+p^{\prime\,2}}{8M_0^2} + \frac{\vec \sigma\cdot\vec p^{\,\prime} \ \vec \sigma\cdot\vec p}{4M_0^2} \,,\\
 \bar u_i \vec\gamma u_j &\approx \vec 0 \ +\ \frac{(\vec p + \vec p^{\,\prime}) + \mathrm i(\vec p-\vec p^{\,\prime})\times\vec\sigma}{2M_0}  \,,\\
 \bar u_i \gamma_5 u_j &\approx 0 \ +\ \frac{\vec\sigma\cdot(\vec p -\vec p^{\,\prime})}{2M_0}  \,,\\
 \bar u_i \gamma^0\gamma_5 u_j &\approx 0 \ +\ \frac{\vec\sigma\cdot(\vec p +\vec p^{\,\prime})}{2M_0} \,,\\
 \bar u_i \vec\gamma\gamma_5 u_j &\approx \vec\sigma + \frac{p^2+p^{\prime2}}{8M_0^2}\vec\sigma + \frac{\vec\sigma\cdot\vec p^{\,\prime} \ \vec\sigma \ \vec\sigma\cdot\vec p}{4M_0^2} \,,\\
 \bar u_i \sigma^{0l} u_j &\approx 0 \ +\ \mathrm i\frac{(p^l - p^{\prime l}) + \mathrm i\epsilon^{lmn}(p^m+p^{\prime m})\sigma^n}{2M_0} \,,\\
 \bar u_i \sigma^{kl} u_j &\approx \epsilon^{klm}\sigma^m \ +\  \epsilon^{klm}\frac{p^2+p^{\prime\,2}}{8M_0^2}\sigma^m - \epsilon^{klm}\frac{\vec\sigma\cdot\vec p^{\,\prime} \ \sigma^m \ \vec\sigma\cdot\vec p}{4M_0^2}\,.
\end{aligned} \end{equation}
If a derivative \(D_\mu\) is contracted with one of the Dirac matrices \(\gamma_5\gamma^\mu\) or \(\sigma^{\mu\nu}\) the matrix element has no order \(\mathcal O (q^0)\) contribution, since \(\bar u_i \gamma_5\gamma^0 u_j=\mathcal O (q^1)\) and \(\bar u_i\sigma^{0\nu}u_j=\mathcal O (q^1)\).
The corresponding term starts in the non-relativistic power counting at least one order higher than in the covariant power counting.
The same argument holds for the product operator \(\epsilon^{\mu\nu\rho\tau}\mathbbm1 D_\rho \otimes \mathbbm1 D_\tau\) sandwiched between Dirac spinors. The leading components from the index combination \(\rho=\tau=0\) gets nullified by the antisymmetric \(\epsilon\)-tensor.


\end{document}